\documentclass[12pt]{emulateapj}
\usepackage{natbib}
\usepackage{longtable}
\usepackage{mathrsfs}
\usepackage{url}

\newcommand{\kms}{\ifmmode {\rm km\ s}^{-1} \else km s$^{-1}$\fi}
\newcommand{\ergs}{\ifmmode {\rm erg\ s}^{-1} \else erg s$^{-1}$\fi}
\newcommand{\ergscm}{\ifmmode {\rm erg\ s}^{-1} \else erg s$^{-1}$ cm$^{-2}$\fi}
\newcommand{\Msun}{\ifmmode {\rm M}_{\odot} \else $M_{\odot}$\fi }
\newcommand{\Lsun}{\ifmmode {\rm L}_{\odot} \else L$_{\odot}$\fi}
\newcommand{\qo}{\ifmmode q_{\rm o} \else $q_{\rm o}$\fi}
\newcommand{\Ho}{\ifmmode H_{\rm o} \else $H_{\rm o}$\fi}
\newcommand{\ho}{\ifmmode h_{\rm o} \else $h_{\rm o}$\fi}

\newcommand{\vFWHM}{\ifmmode v_{\mbox{\tiny FWHM}} \else
                    $v_{\mbox{\tiny FWHM}}$\fi}
\newcommand{\CCF}{\ifmmode F_{\it CCF} \else $F_{\it CCF}$\fi}
\newcommand{\ACF}{\ifmmode F_{\it ACF} \else $F_{\it ACF}$\fi}
\newcommand{\Halpha}{\ifmmode {\rm H}\alpha \else H$\alpha$\fi}
\newcommand{\Hbeta}{\ifmmode {\rm H}\beta \else H$\beta$\fi}
\newcommand{\Hgamma}{\ifmmode {\rm H}\gamma \else H$\gamma$\fi}
\newcommand{\Hdelta}{\ifmmode {\rm H}\delta \else H$\delta$\fi}
\newcommand{\Lya}{\ifmmode {\rm Ly}\alpha \else Ly$\alpha$\fi}
\newcommand{\Lyb}{\ifmmode {\rm Ly}\beta \else Ly$\beta$\fi}
\newcommand{\HeI}{\ifmmode {\rm He}\,{\sc i}\,\lambda5876 \else 
	          He\,{\sc i}\,$\lambda5876$\fi}
\newcommand{\HeII}{\ifmmode {\rm He}\,{\sc ii}\,\lambda4686 \else 
	           He\,{\sc ii}\,$\lambda4686$\fi}

\newcommand{\ciii}{\ifmmode {\rm C}\,{\sc iii} \else C\,{\sc iii}\fi}

\newcommand{\caii}{Ca\,{\sc ii}}

\newcommand{\mbh}{$M_{\rm BH}$\ }
\newcommand{\sigstar}{$\sigma_{*}$\ }

\newcommand{\rblr}{$R_{\rm BLR}$\ }
\newcommand{\msigma}{$M_{\rm BH}$--$\sigma_{*}$\ }

\shorttitle{$M_{\rm BH}$--$\sigma_*$ Relation}
\shortauthors{Grier et al.}

\begin{document}

\title{Stellar Velocity Dispersion Measurements in High-Luminosity
Quasar Hosts and Implications for the AGN Black Hole Mass Scale}

\author{C.~J.~Grier\altaffilmark{1},
P.~Martini\altaffilmark{1,2,3},
L.~C.~Watson\altaffilmark{4},
B.~M.~Peterson\altaffilmark{1,2},
M.~C.~Bentz\altaffilmark{5}, 
K.~M.~Dasyra\altaffilmark{6}, 
M.~Dietrich\altaffilmark{7}, 
L.~Ferrarese\altaffilmark{8}, 
R.~W.~Pogge\altaffilmark{1,2}, \&
Y. Zu\altaffilmark{1}}
\altaffiltext{1}{Department of Astronomy, The Ohio State University,
140 W 18th Ave, Columbus, OH 43210, USA} 
\altaffiltext{2}{Center for Cosmology and AstroParticle Physics, The 
Ohio State University, Columbus, OH 43210, USA}
\altaffiltext{3}{Visiting Astronomer, North American ALMA Science
 Center and University of Virginia, Charlottesville, VA 22903, USA}
\altaffiltext{4}{Harvard-Smithsonian Center for Astrophysics, 
60 Garden Street, Cambridge, MA 02138, USA}
\altaffiltext{5}{Department of Physics 
and Astronomy, Georgia State University, Atlanta, GA 30303, USA} 
\altaffiltext{6}{Observatoire de Paris, LERMA (CNRS:UMR8112), 
61 Av. de l'Observatoire, F-75014, Paris, France}
\altaffiltext{7}{Department of Physics and Astronomy, Ohio University, 
Athens, OH 45601, USA}
\altaffiltext{8}{Herzberg Institute of Astrophysics, National Research 
Council of Canada} 

\begin{abstract}
  We present new stellar velocity dispersion measurements for four
  luminous quasars with the NIFS instrument and the ALTAIR laser guide
  star adaptive optics system on the Gemini North 8-m
  telescope. Stellar velocity dispersion measurements and measurements
  of the supermassive black hole masses in luminous quasars are
  necessary to investigate the coevolution of black holes and
  galaxies, trace the details of accretion, and probe the nature of
  feedback. We find that higher-luminosity quasars with higher-mass
  black holes are not offset with respect to the \msigma relation
  exhibited by lower-luminosity AGNs with lower-mass black holes, nor
  do we see correlations with galaxy morphology. As part of this
  analysis, we have recalculated the virial products for the entire
  sample of reverberation-mapped AGNs and used these data to
  redetermine the mean virial factor $\langle f \rangle$ that places
  the reverberation data on the quiescent \msigma relation. With our
  updated measurements and new additions to the AGN sample, we obtain
  $\langle f \rangle$~=~4.31~$\pm$~1.05, which is slightly lower than,
  but consistent with, most previous determinations.
\end{abstract}

\section{INTRODUCTION}
Over the past couple of decades, both observational and analytical
work have suggested a physical connection between the formation and
growth of galaxies and the growth of their central black holes. For
example, the comoving emissivity of active galactic nuclei (AGNs) and
the cosmic star formation rate have both similarly declined since $z
\sim 1$ (\citealt{Boyle98}; \citealt{Franceschini99};
\citealt{Merloni04}; \citealt{Silverman08}), which might imply a link
between star formation and AGN activity. In addition, luminous AGNs
are more often found in massive early-type galaxies with young stellar
populations (e.g., \citealt{Sanders88}; \citealt{Kauffmann03};
\citealt{Veilleux09}). Further support of a black hole--galaxy
connection comes in the form of a number of correlations between
properties of the host galaxies and the masses of their central black
holes (BHs). A key relationship is between black hole mass ($M_{\rm
  BH}$) and bulge stellar velocity dispersion ($\sigma_*$), observed
in both quiescent (\citealt{Ferrarese00}; \citealt{Gebhardt00a};
\citealt{Tremaine02}; \citealt{Gultekin09}; \citealt{McConnell11};
\citealt{McConnell13}) and active galaxies (\citealt{Gebhardt00b};
\citealt{Ferrarese01}; \citealt{Nelson04}; \citealt{Onken04};
\citealt{Dasyra07}; \citealt{Woo10}; \citealt{Graham11};
\citealt{Park12}). This relation was first predicted by \cite{Silk98}
and \cite{Fabian99} and has been explained by various analytic models
(e.g., \citealt{King03}; \citealt{King05}; \citealt{Murray05}) as well
as recovered in numerical simulations of evolving and interacting
galaxies \citep[e.g.,][]{Dimatteo05, Dimatteo08}. The
\msigma relationship can be used to infer \mbh in large samples of
galaxies. This allows for the exploration of the BH mass function on
much larger scales (e.g., \citealt{Yu02}) and thus helps investigate
the role of BHs in galaxy formation and evolution processes.

Direct \mbh measurements are made with stellar kinematics and gas
dynamics, although these methods require good spatial resolution and
are presently only feasible for nearby galaxies. AGNs, however, offer
the most robust tracer of the evolution of the BH population over much
of the history of the universe. Under the assumption that the motion
of the gas in the broad line region (BLR) of AGNs is dominated by the
gravitational influence of the black hole, one can use the virial
relation $M_{\rm BH} = (f R_{\rm BLR} \Delta V^2)/G$ to obtain $M_{\rm
  BH}$, where \rblr is the average radius of the emitting gas in the
BLR, usually either determined with reverberation mapping (e.g.,
\citealt{Peterson04}) or estimated with the radius-luminosity relation
(e.g., \citealt{Bentz09a, Bentz13}), $\Delta V$ is the
velocity dispersion of the gas, deduced from the width of the emission
line, and $f$ is a dimensionless factor that accounts for the unknown
geometry and orientation of the BLR and may be different for each AGN.

With current technology, we are unable to directly observe the
structure of the BLR, as it is unresolvable even with the largest
telescopes, so the true value of $f$ for each object is unknown. This
has contributed significantly to the uncertainties in \mbh
measurements using BLR emission lines. Recent reverberation mapping
(RM) efforts have begun to reveal more information about the actual
structure of the BLR and the value of $f$ in some objects (e.g.,
\citealt{Bentz10b}; \citealt{Brewer11}; \citealt{Pancoast12};
\citealt{Grier13a}). However, limited data for most AGNs requires the
use of an average virial factor $\langle f \rangle$ to estimate
$M_{\rm BH}$. Currently, $\langle f \rangle$ is calculated with the
assumption that AGNs follow the same \msigma relation as quiescent
galaxies (\citealt{Onken04}; \citealt{Woo10}; \citealt{Graham11};
\citealt{Park12}; \citealt{Woo13}). Most estimates of $\langle f
\rangle$ are somewhat larger than $\sim 5$; \cite{Onken04} find
$\langle f \rangle$ = 5.5 $\pm$ 1.8, \cite{Woo10} find $\langle f
\rangle$ = 5.2 $\pm$ 1.2, and more recently, analysis by \cite{Park12}
and \cite{Woo13} both yield $\langle f \rangle$ = 5.1. \cite{Graham11}
obtain a slightly lower value, $\langle f \rangle$ =
3.8$^{+0.7}_{-0.6}$.

The difference between slopes and virial factors among studies using
similar regression methods (whether \mbh is considered the independent
or dependent variable) arise when different galaxy samples are used to
determine these two quantities, which may suggest a morphological
dependence or selection bias in the relation. In fact, recent studies
do report a morphological dependence in the quiescent \msigma
relation, such that there are systematic differences in the relation
for early-type (higher-mass) and late-type (lower-mass) galaxies
(e.g., \citealt{Greene10}; \citealt{McConnell13}). Others have found
that barred galaxies lie systematically below the \msigma relation of
normal unbarred galaxies \citep[e.g.,][]{Graham08a, Graham08b,
  Graham09}, and still others have found deviations in both the slope
and intercept for galaxies hosting pseudobulges (e.g., \citealt{Hu08};
\citealt{Gadotti09}; \citealt{Kormendy11}). The idea of a
non-universal \msigma relation has been supported by theoretical work
as well (e.g., \citealt{King10}; \citealt{Zubovas12}), which has also
suggested that the relation may depend on environment.

Morphological deviations from a single \msigma relation have also been
claimed in AGNs (e.g., \citealt{Graham09}; \citealt{Mathur12}), and
there has been some question as to whether or not objects at the
high-mass/high-\sigstar end of the relation follow a different slope
(e.g., \citealt{Dasyra07}; \citealt{Watson08}). For example, four out
of the six objects with \mbh above 10$^8$ \Msun \ included in the
study of \cite{Watson08} lie significantly above the relation. The
appearance of outliers could be due to systematic errors in \sigstar
or \mbh measurements, or simply a fluke due to small number
statistics. Alternatively, \cite{Lauer07} suggest that offsets at the
high-mass end may be due to a selection bias. Specifically, when a
sample is selected based on AGN properties, one is more likely to find
a high-mass BH in a lower-mass galaxy (based on a BH--host galaxy
correlation) because high mass galaxies are rare and there is
intrinsic scatter in BH--host galaxy correlations.

An important step in evaluating the \msigma relation and any possible
deviations from it is to obtain secure \sigstar and \mbh measurements
in AGNs that sample the entire mass range of the relation. While the
high-mass end of the quiescent \msigma relation is relatively
well-populated to beyond 10$^9$ \Msun \ (\citealt{McConnell13}), the
current sample of AGNs used to calculate $\langle f \rangle$ still
contains just three or four objects with \mbh above 10$^8$ \Msun \
(\citealt{Graham11}; \citealt{Park12}; \citealt{Woo13}). More
measurements for luminous AGNs are needed to measure the high-mass end
of the AGN \msigma relation. However, accurate \sigstar measurements
for high-luminosity AGNs are difficult to obtain because the AGN light
overpowers the light from the host. Moreover, more luminous AGNs are
relatively scarce and thus typically found at large distances, so the
host galaxy has a small angular size and is easily lost in the glare
of the AGN. It is only in the past few years that high-precision
measurements in very luminous objects have been obtained on account of
the availability of adaptive optics (AO) and integral field
spectrographs (IFUs) such as Gemini North's Near-Infrared Integral
Field Spectrometer (NIFS) combined with the Gemini North laser guide
star AO system, ALTAIR. \cite{Watson08} used NIFS+ALTAIR to measure
\sigstar for PG\,1426+015 with much higher precision than previous
measurements for high-luminosity quasars. This success prompted us to
undertake additional observations of quasars at the high-mass end of
the \msigma relation. In this paper we present the results of our NIFS
observations of eight high-luminosity quasars. We successfully
measured \sigstar in four objects and use these results to improve the
population of the \msigma relation at the high-mass end. We also
recalculate virial products for the entire AGN sample with updated
time lag measurements to re-derive $\langle f \rangle$, calibrate
black hole masses in AGNs, and reexamine the AGN \msigma relation. In
this work we adopt a cosmological model of $\Omega_{m}=0.3$,
$\Omega_{\Lambda}=0.70$, and $H_0 = 70$ km sec$^{-1}$ Mpc$^{-1}$.

\section{OBSERVATIONS AND DATA ANALYSIS}

\subsection{NIFS/ALTAIR Observations}
Observations of eight quasars were carried out at the Gemini North
telescope in 2008 and 2010 under the programs GN-2008B-Q-28,
GN-2010A-Q-11, and GN-2010B-Q-24. We chose our sample from the
database of objects with RM-based black hole mass measurements from
\cite{Peterson04} with \mbh $>$ 10$^8$ \Msun. Basic information on our
targets is given in Table \ref{Table:tbl1}. We used NIFS in
conjunction with the ALTAIR laser guide star AO system to carry out
our observations. NIFS has a 3\arcsec $\times$ 3\arcsec \ field of
view that is divided into 29 individual spectroscopic slices, with a
spectral resolution $R$ = $\lambda/\Delta \lambda$ $\approx$ 5290 in
both the $H$ and $K$ bands. With the AO correction, NIFS yields a
spatial resolution on the order of 0.1\arcsec. There are several
strong stellar absorption lines that fall within the wavelength range
of the $H$ band, which has a central wavelength of 1.65 $\mu$m and
covers from about 1.49 $\mu$m to 1.80 $\mu$m, so we observed seven of
our targets with the $H$ band filter. We list the most prominent
stellar absorption features in this wavelength region in Table
\ref{Table:tbl3}.  We observed our eighth object, PG\,1700+518, in the
$K$ band due to its higher redshift. The $K$ filter on NIFS covers
from about 1.99 $\mu$m to 2.40 $\mu$m.

We estimated the integration time for each object with $HST$ ACS or
WFPC2 images of the sources from \cite{Bentz09a}. To simulate the
data we would obtain from NIFS, we measured the flux within a 3\arcsec
x 3\arcsec \ aperture, except for a central circle of diameter
0.2\arcsec. We estimated the exposure time for each object based on its
brightness relative to PG 1426+015, for which \cite{Watson08} obtained
a host-galaxy signal-to-noise ratio ($S/N$) $\sim$ 200 in about two
hours of on-source integration. With both Poisson and
background-limited trials, we estimated the integration time required
for each object to obtain a $S/N$ $\sim$ 200. Table \ref{Table:tbl2}
gives details of the observations, most notably the on-source
integration time for each object. Reconstructed images from the IFU
spectra of all eight targets are shown in Figure 1. We
observed telluric standard stars (usually A0V) once every 1.5 hours
for the purpose of telluric corrections.

\begin{figure}
\begin{center}
\epsscale{1.2}
\plotone{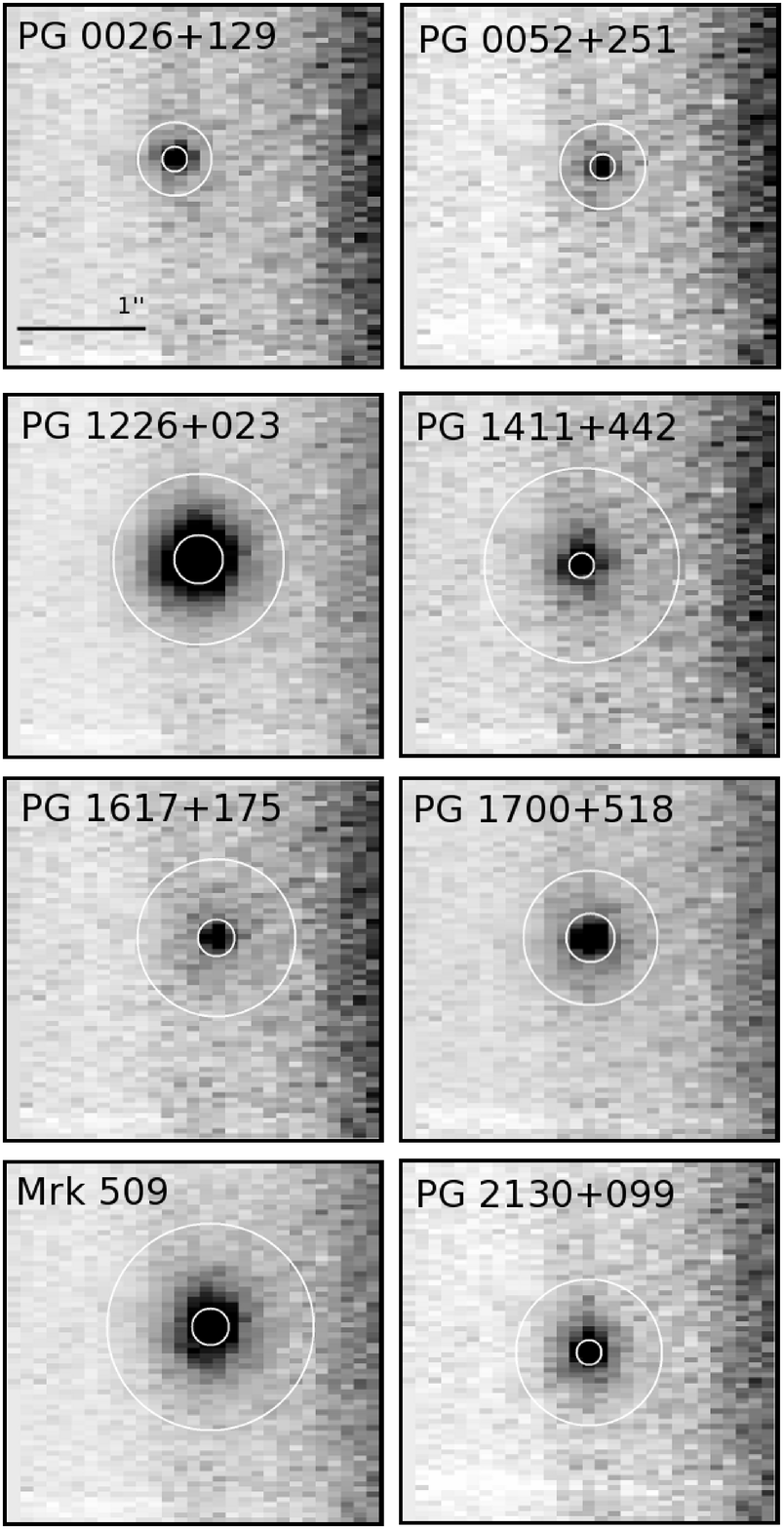}
\caption{Raw reconstructed images for each object. The white circles
  denote the extraction annuli used for each object, given in Table
  \ref{Table:tbl4}. The field of view of each panel is
  3\arcsec$\times$3\arcsec. }
\label{fig:mosaic}
\end{center}
\end{figure}

\subsection{Data Reduction}
Data were processed through the standard NIFS
pipeline\footnote{http://www.gemini.edu/sciops/instruments/nifs} from
the Gemini IRAF\footnote{IRAF (\citealt{Tody86}) is distributed by the
  National Optical Astronomy Observatories, which are operated by the
  Association of Universities for Research in Astronomy, Inc., under
  cooperative agreement with the National Science Foundation.}
package. Our reductions deviated from the standard pipeline tasks in
only two ways. First, we found that the original sky frames did not
adequately remove sky lines from our spectra. To remedy this, we
manually scaled the individual sky spectra to obtain better sky
subtraction in each individual object frame. Second, to remove stellar
absorption lines in our telluric spectra, we used methods described by
\cite{Vacca03} and applied in the IDL-based code \emph{xtellcor}. This
code uses a theoretical model of Vega to remove the hydrogen features
in our telluric standard spectra, and is specifically written for use
with A0V stars.

To separate the host-galaxy spectra from the AGNs, we extract the
spectrum from an annulus that excludes the quasar-dominated
nucleus. The use of AO in these observations allowed us in most cases
to confine at least the core of the quasar flux to the very central
pixels of the image. Generally, the AO-assisted seeing was on the
order of $0.1-0.2$\arcsec, so we used either a $0.2$\arcsec,
0.3\arcsec, or $0.4$\arcsec \ inner radius ($R_{\rm inner}$) to
isolate the quasar component. The outer radius for each extraction
annulus ($R_{\rm outer}$) was chosen to include as much host galaxy
light as possible while minimizing the amount of noise contributed by
the sky. For most of our objects, we evaluated this by eye and chose
windows that minimized noise. However, in the targets where we could
see identifiable galactic absorption lines, we chose $R_{\rm inner}$
and $R_{\rm outer}$ to obtain the highest equivalent width
measurements in the visible absorption lines. $R_{\rm inner}$ and
$R_{\rm outer}$ for each object are listed in Table \ref{Table:tbl4}
and are shown on the reconstructed images in Figure
1. The total galaxy+quasar spectra for all eight
objects, extracted from within a radius of $R_{\rm outer}$, are shown
in the top panels of Figure~2.

Although the use of the ALTAIR AO system helps confine the nuclear
light to the central few pixels of the image, the AO-corrected PSF
still has nuclear light in its wings. Typical Strehl ratios for the
ALTAIR AO system are on the order of 0.1 to 0.3 in the $H$ band
(\citealt{Christou10}). For a typical Strehl ratio of 0.2,
\cite{Christou10} report a 50\% encircled PSF energy radius of about
0.4\arcsec; i.e., half of the quasar light falls in a radius outside
0.4\arcsec. Thus, there is still significant quasar contamination in
the host-galaxy spectra. To further remove the quasar emission, we
scaled and subtracted the nuclear spectrum from our annulus in each
target. The scaling was done empirically --- we chose a scale factor
that best eliminated the most prominent quasar emission lines seen in
the spectra. Our final nucleus-subtracted, observed-frame spectra of
all eight objects are shown in the bottom panels of Figure~2. Note
that while the spectrograph coverage in the $H$ band extends from
about 1.48 $\mu$m \ to 1.8 $\mu$m, in some cases the telluric
contamination was sufficiently significant that we cropped the spectra
before using them. The wavelength coverage of the $H$-band spectra
shown in Figure~2 is not uniform for this reason.

\begin{figure}
\figurenum{2}
\epsscale{1.2}
\plotone{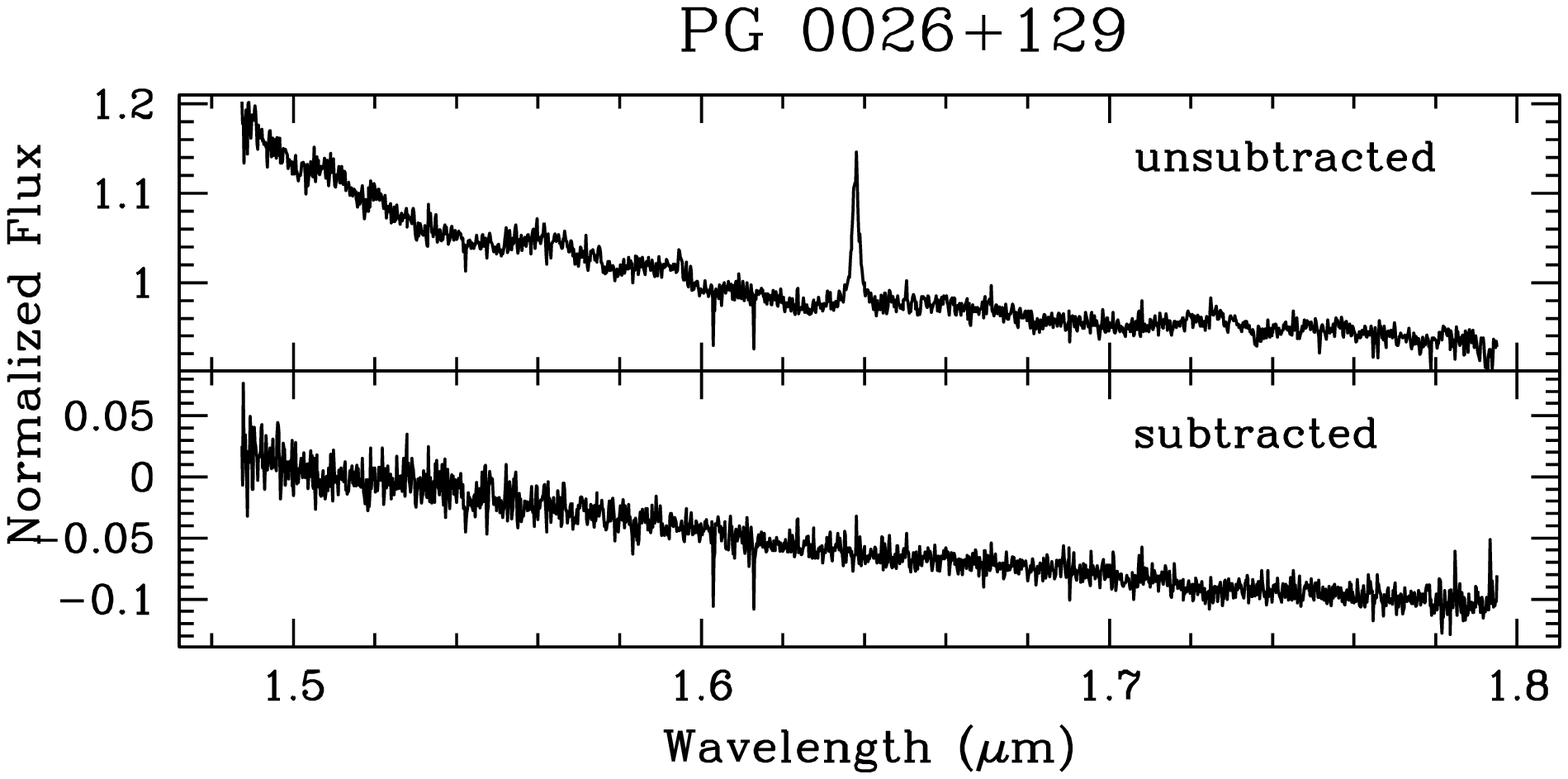}
\plotone{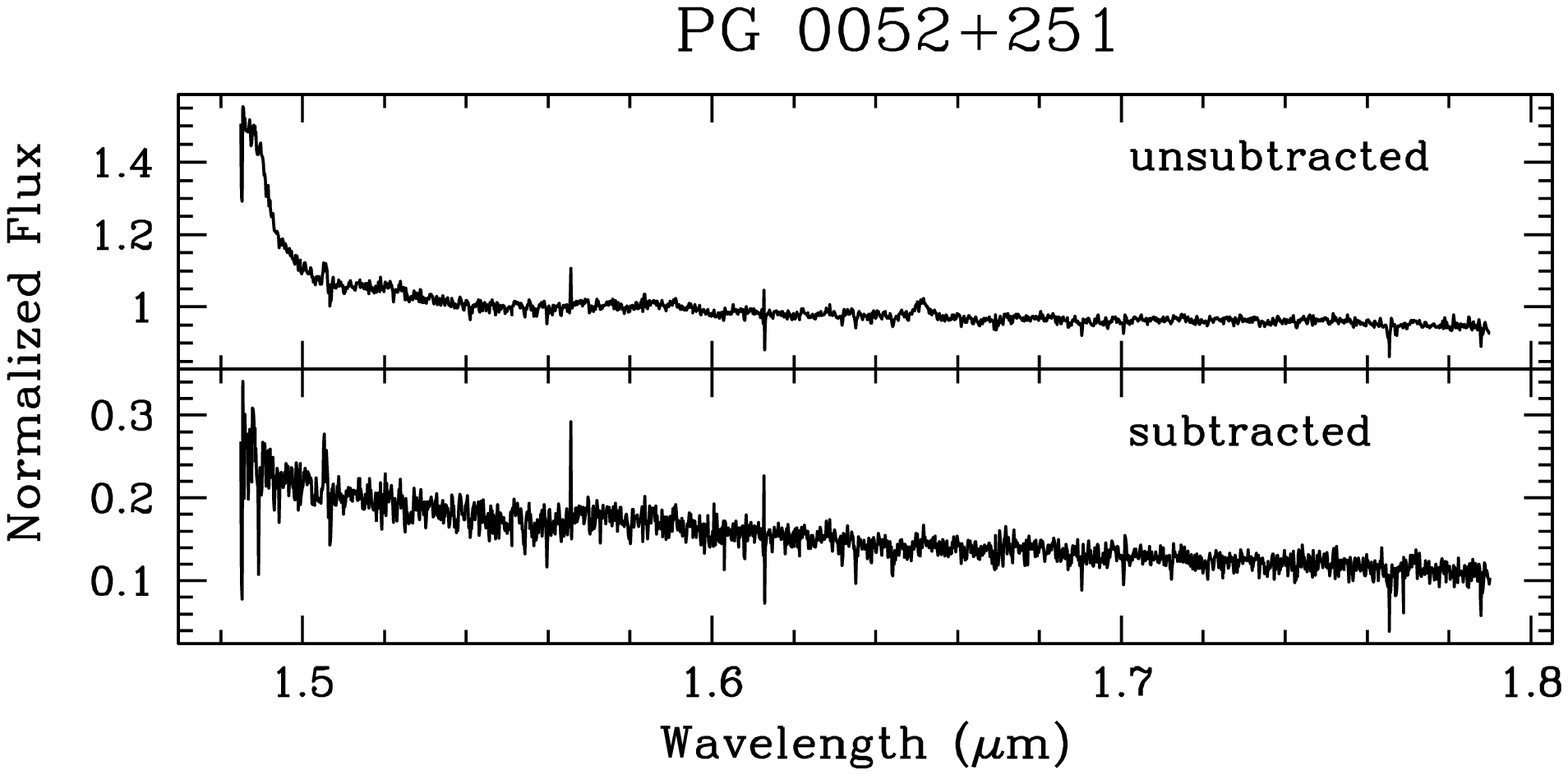}
\plotone{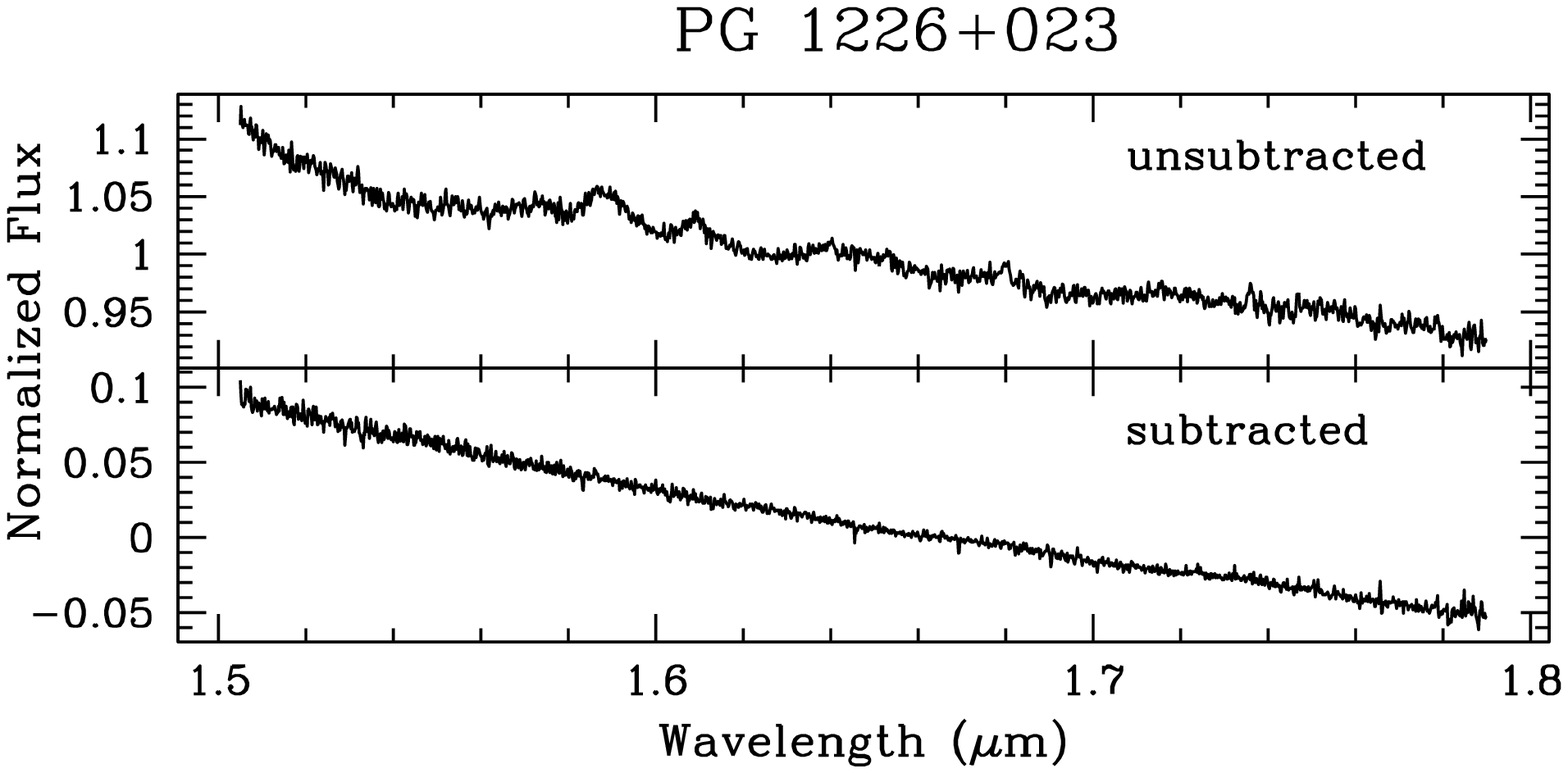}
\plotone{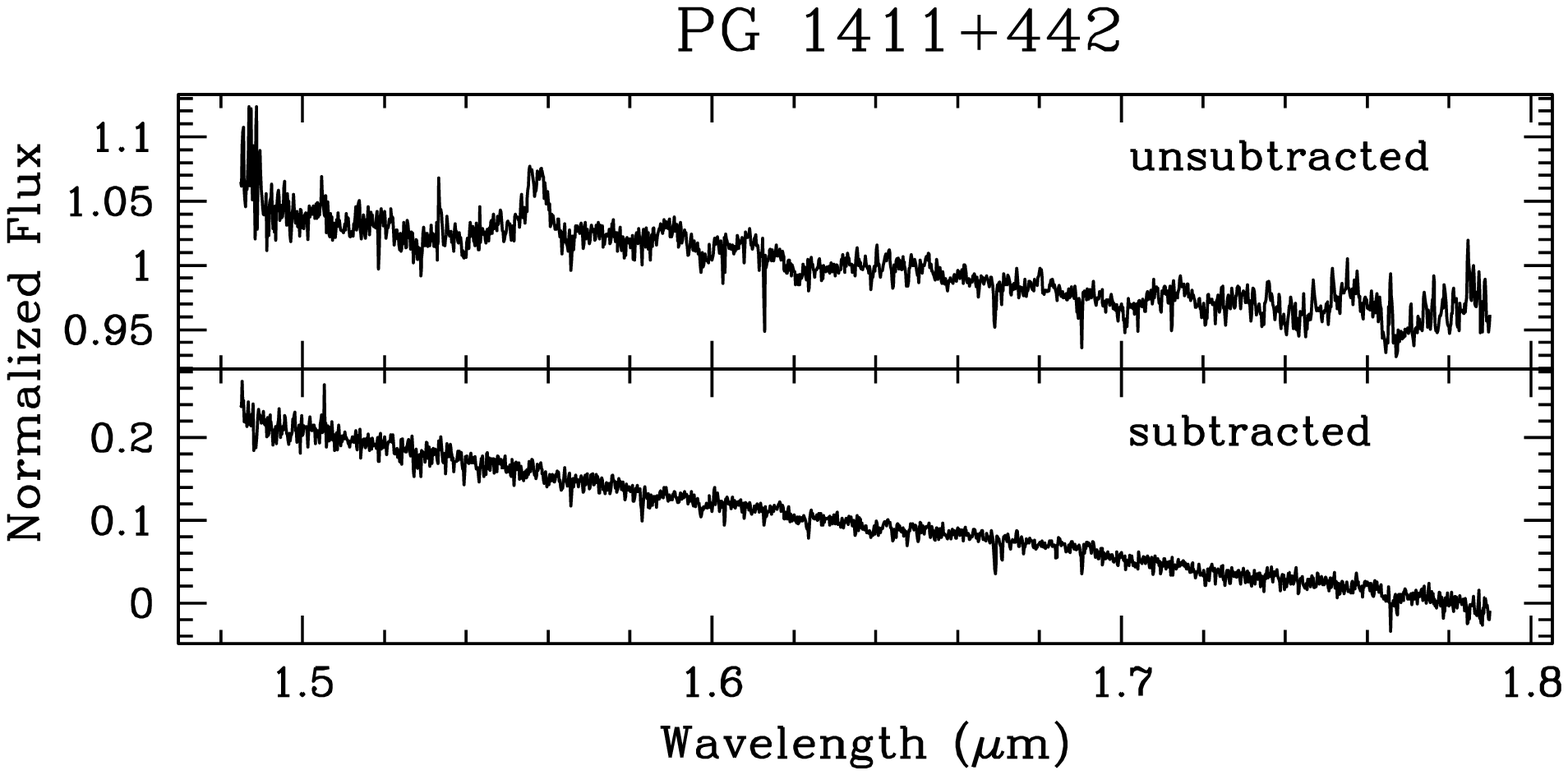}
\caption{Observed-frame spectra of all eight targets. The top panels
  show the total, unsubtracted spectrum, and the bottom panels show
  the host galaxy spectrum after the nucleus was subtracted off. The
  fluxes are in units of flux per unit wavelength, and have been
  normalized to the mean of the unsubtracted spectrum for each object. }
\end{figure}

\begin{figure}
\figurenum{2}
\epsscale{1.2}
\plotone{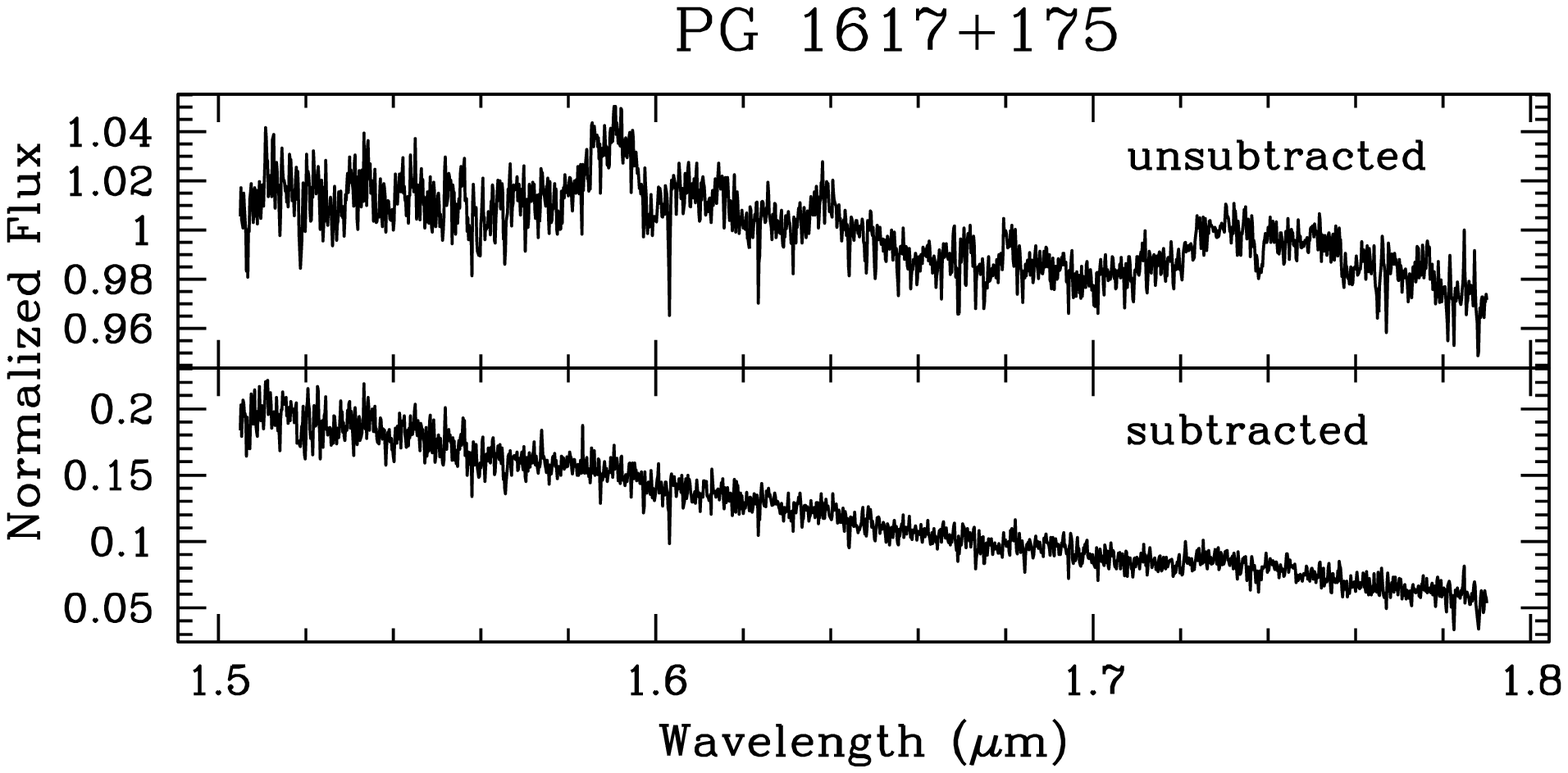}
\plotone{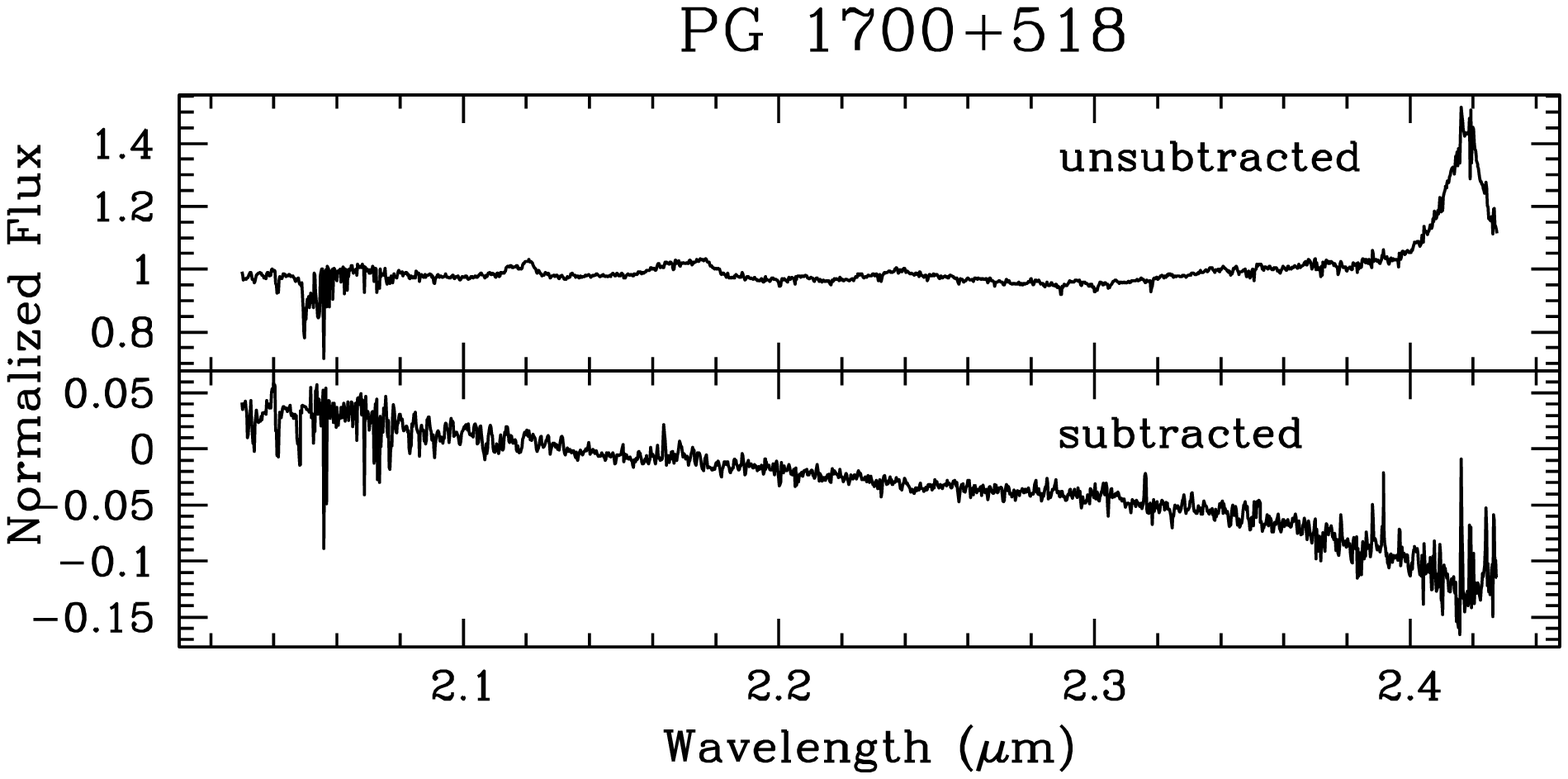}
\plotone{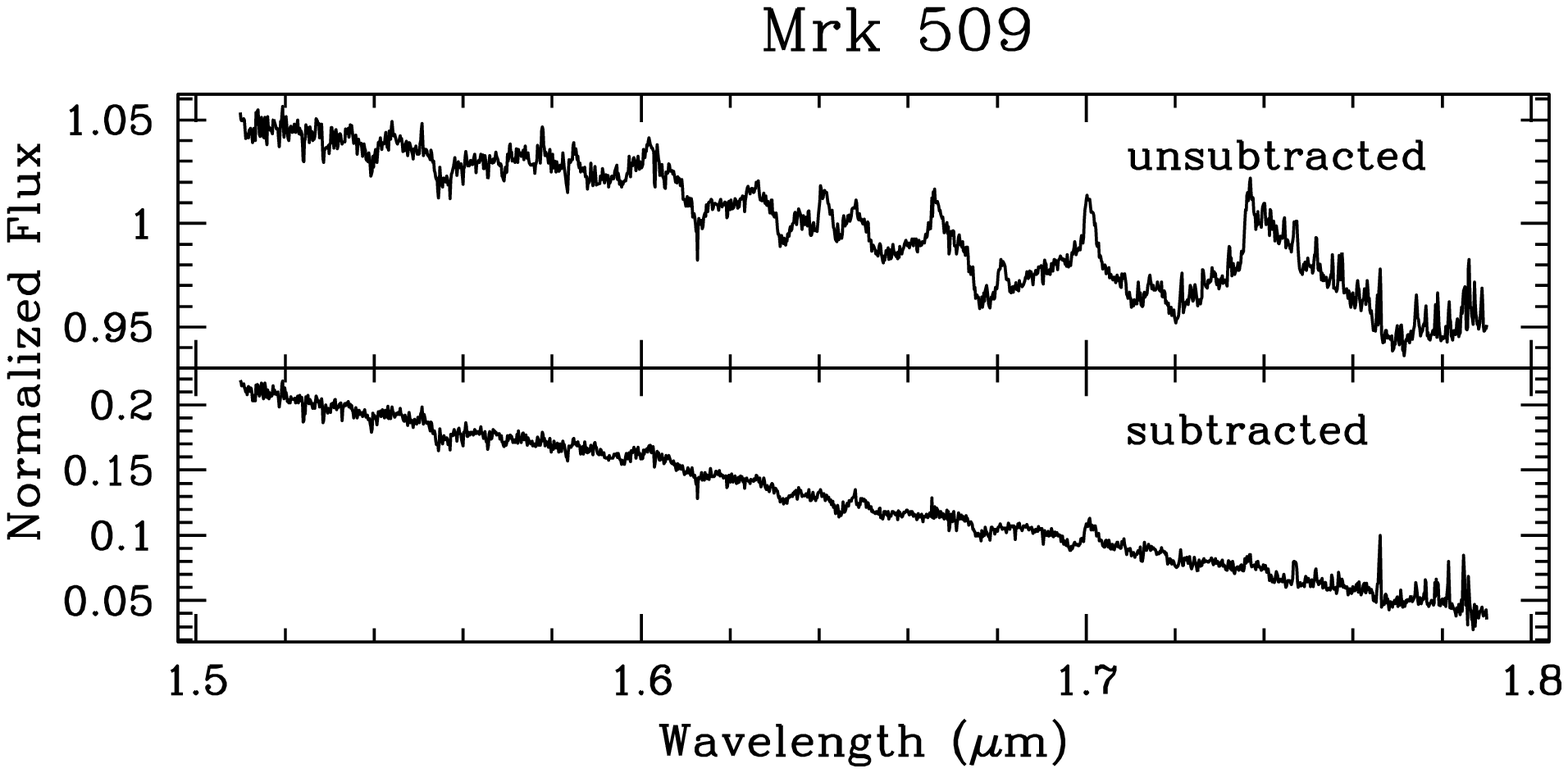}
\plotone{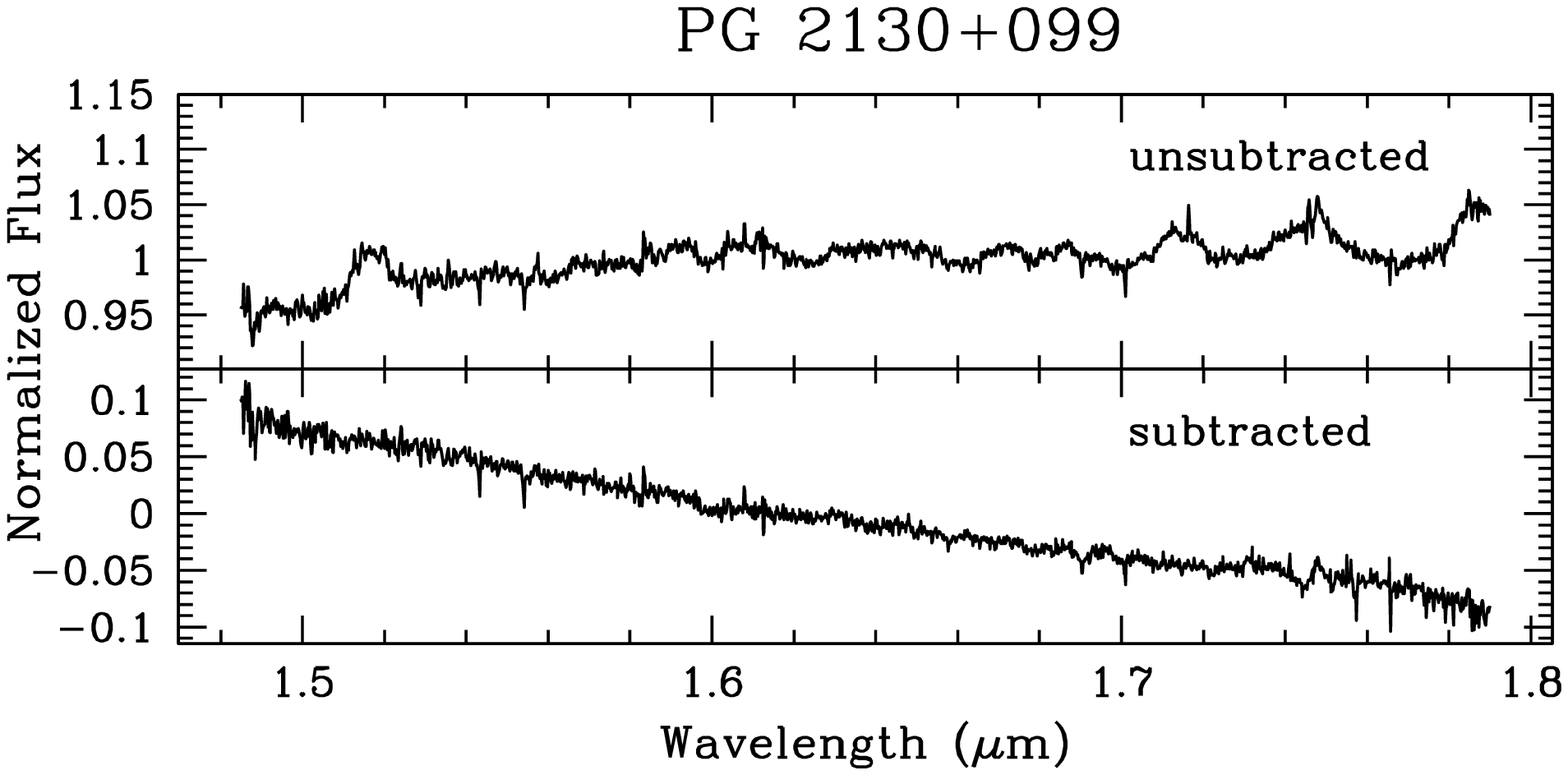}
\caption{{\it Continued.}}
\end{figure}
\subsection{Stellar Velocity Dispersion Measurements}
We used the penalized pixel fitting method (pPXF) of
\cite{Cappellari04} to measure $\sigma_*$. This method convolves a
stellar template spectrum and a line-of-sight velocity distribution to
model the host galaxy. The best line-of-sight velocity distribution is
calculated in the pPXF code with a $\chi^{2}$ minimization
technique. The velocity templates used to make our measurements were
obtained by \cite{Watson08}, and include stars of four different
spectral classes: K0\,III, K5\,III, M1\,III, and M5\,Ia. The K5\,III,
M1\,III, and M5\,Ia templates all resulted in somewhat similar fits in
each spectrum --- the reduced $\chi^2$ values of the fit in these
three cases were always very close to one another, with the K5\,III
template usually a slightly better fit than the other two. The K0\,III
template provided a poor fit to the CO(3-0) absorption line in all
objects and was therefore not used in our analysis.

There are several different factors that affect the uncertainty in
$\sigma_*$. First, no single stellar template is expected to be a
perfect match to the host-galaxy stellar absorption features. We
therefore adopt the average \sigstar value from the three templates as
our estimate, and fold the differences into our uncertainties. One
exception to this is Mrk 509, for which the K5\,III template provided
a significantly better fit. We take the standard deviation in \sigstar
reported by the pPXF software among the three stellar templates as
representative of the template mismatch uncertainty. We also consider
the location of $\Delta \chi^2$ = 1 in the fitting a component of our
uncertainties. To do this, we allowed \sigstar to vary but held the
rest of the parameters fixed at their best-fit values. We identified
the value of \sigstar at which the $\chi^2$ had changed by $\pm$ 1
from the best-fit value, and took the average difference between these
two values and our best-fit value as our uncertainty. Because our
measurements for PG\,1411+442 and PG\,1617+175 were made with spectra
that are significantly noisier than the other two spectra, we also
include a signal-to-noise ($S/N$) component in our uncertainties for
those two objects. To estimate this component, we degraded our two
best spectra (Mrk 509 and PG\,2130+099) to match the $S/N$ in
PG\,1617+175 and PG\,1411+442. We then re-measured \sigstar in the
degraded spectra and took the deviation from our original measurements
to represent the $S/N$ component of the uncertainties. For most of the
templates, the \sigstar measurements tended to be overestimated in the
degraded spectra by about 20 \kms. We combine the 1$\sigma$
uncertainties, the template mismatch uncertainties, and any $S/N$
component in quadrature and adopt these as our formal uncertainties.

\begin{figure}
\figurenum{3}
\begin{center}
\epsscale{1.2} 
\plotone{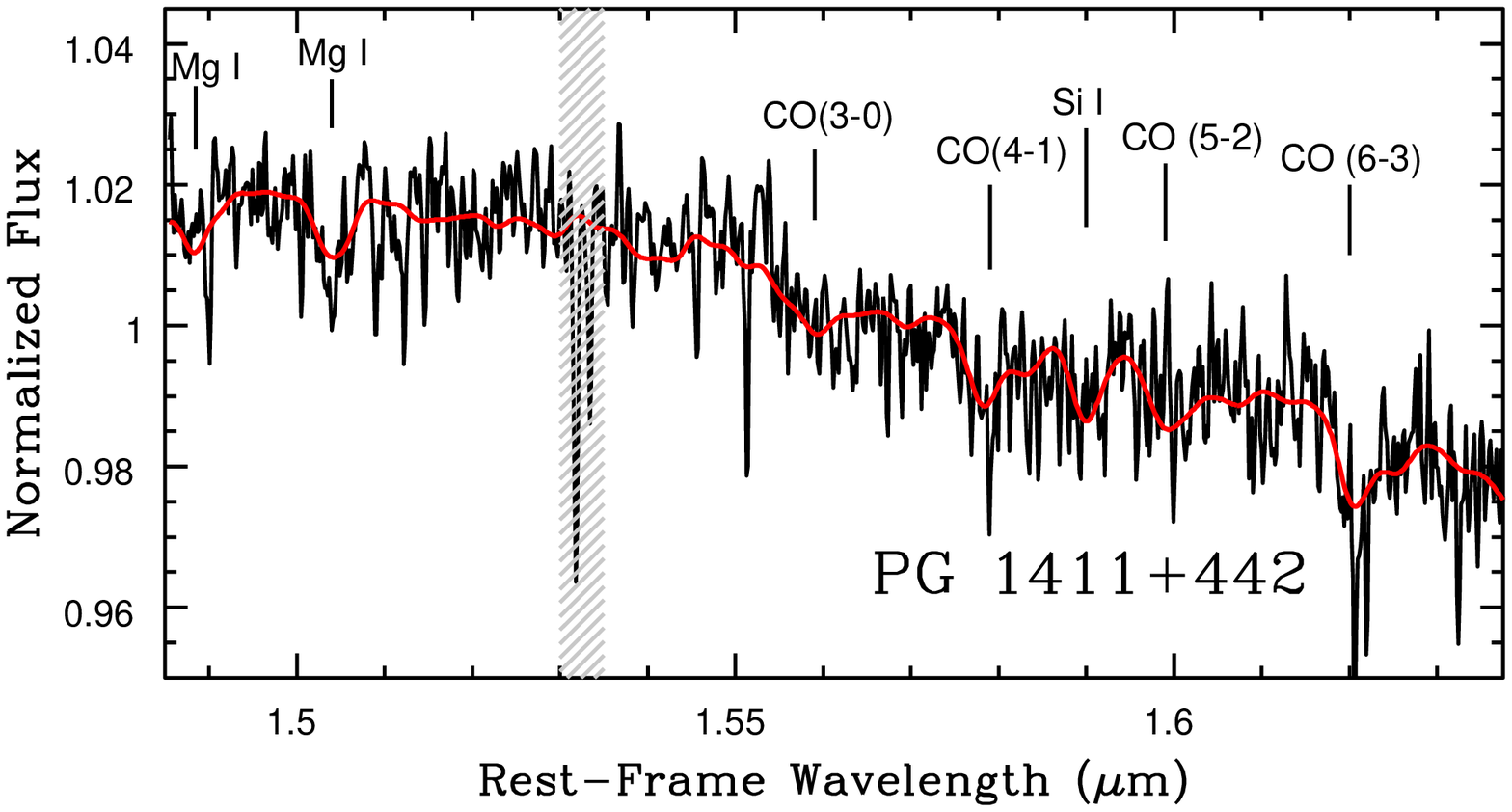}
\plotone{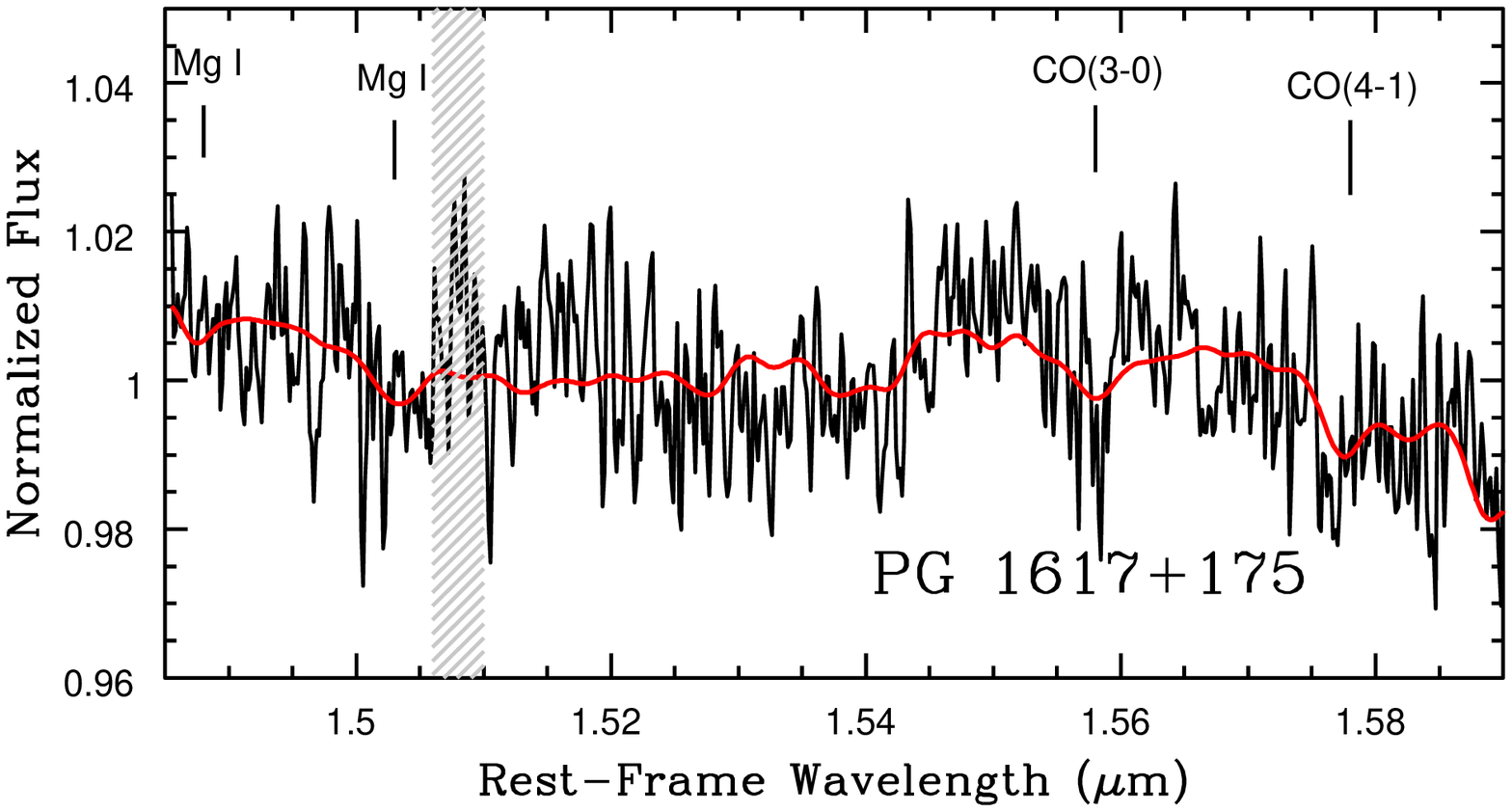}
\plotone{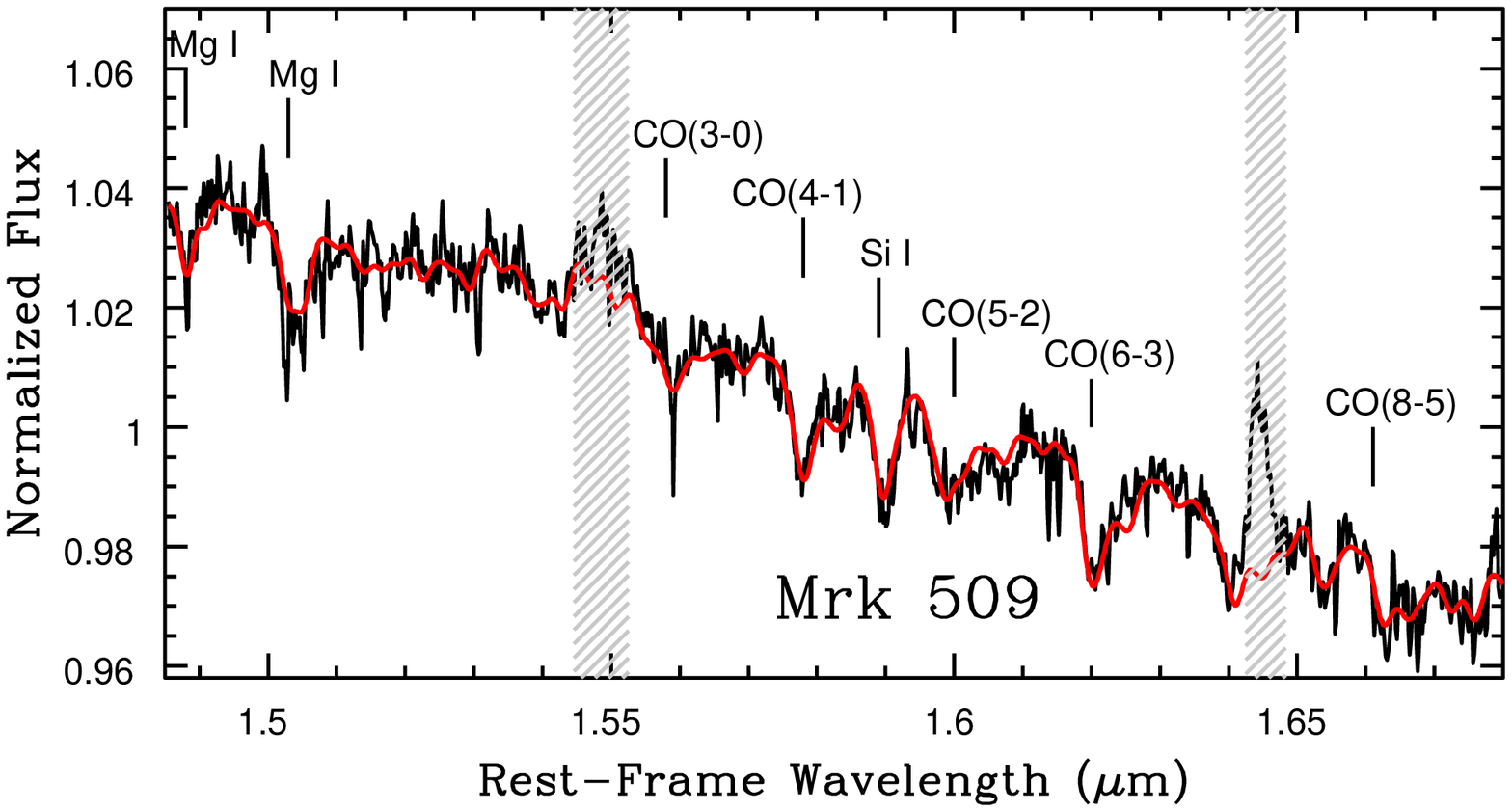}
\plotone{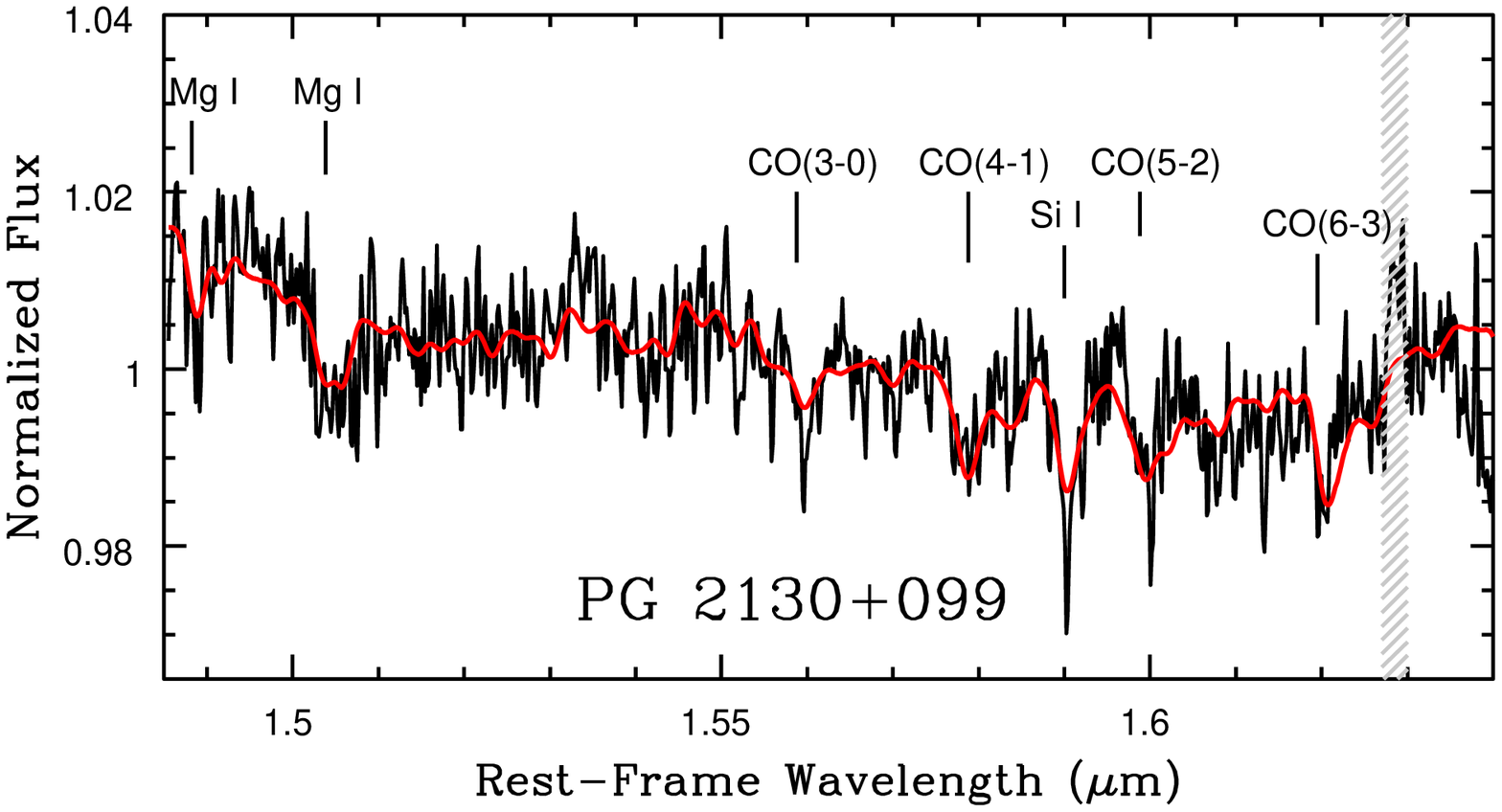}
\caption{Normalized rest-frame spectra of objects in which we measured
  \sigstar successfully. The red lines are our best-fit models, with
  the K5\,III stellar template for PG\,1411+442, Mrk 509, and
  PG\,2130+099, and the M5\,Ia template for PG\,1617+175. The shaded
  gray areas mark areas excluded from the fit due to telluric and/or
  quasar light contamination.}
\label{fig:fits}
\end{center}
\end{figure}

We measured \sigstar from the spectra of four of our objects: Mrk 509,
PG\,1411+442, PG\,1617+175, and PG\,2130+099, and these values are
listed in Table \ref{Table:tbl4}. The normalized galaxy spectra and
the best-fit broadened stellar templates for these four objects are
shown in Figure~3. Our measurement for PG\,2130+099 of 147 $\pm$ 17 km
s$^{-1}$ constitutes a significant improvement in precision over its
previous measurement of 172 $\pm$ 46 km s$^{-1}$ (\citealt{Dasyra07}),
while Mrk 509 and PG\,1411+442 have no previous measurements reported
in the literature. Our measurement for PG\,1617+175 of 201 $\pm$ 37
km$^{-1}$ is slightly more precise but consistent with measurements
made by \cite{Dasyra07}, who report \sigstar = 183 $\pm$ 47 km
s$^{-1}$. In the other four objects, we were unable to identify any
absorption lines in our nucleus-subtracted spectra and were unable to
fit stellar templates and recover $\sigma_*$. We believe this was
caused mainly by the overwhelming strength of the quasar emission in
these objects and a lack of strong absorption lines within the
observed wavelength range. These four objects are also the most
distant of the eight we observed. In an attempt to minimize the quasar
contamination, we experimented with different radii for both the inner
and outer regions in these objects, but in all cases were unable to
remove the quasar contamination enough to see absorption in the host
galaxy spectrum. Even when subtracting off a scaled nuclear spectrum
to eliminate the emission lines, we see only noise (with residual sky
and telluric contamination; see Figure~2) in the spectra of objects in
which we were unable to measure $\sigma_*$.

Because we optimized our extraction radii to obtain the best
host-galaxy-to-quasar ratio, our \sigstar measurements were made
within different effective physical apertures for each target. The
measurements for Mrk 509, PG\,1411+442, and PG\,1617+175 were made
within an effective physical aperture of 1.25 kpc, 3.2 kpc, and 3.3
kpc respectively. The effective radii ($r_{\rm e}$) for the spheroid
components of these three objects from the surface decompositions of
\cite{Bentz09a} are 1.85 kpc, 5.05 kpc, and 3.3 kpc, so our apertures
for these three objects lie between, or very close to, the commonly
quoted aperture sizes of $r_{\rm e}$ and $r_{\rm e}/8$. However, the
measurement for PG\,2130+099 was made within 1.64 kpc, which is 4.3
times $r_{\rm e}$ for this object. We apply the relation derived by
\cite{Jorgensen95} to determine the velocity dispersion within the
effective radius ($\sigma_{*,\rm e}$) for all four objects. These
values are listed in Table \ref{Table:tbl4} for comparison with our
measurements. For Mrk 509, PG\,1411+442, and PG\,1617+175, there is
very little difference between our measured \sigstar and
$\sigma_{*,\rm e}$. The value of $\sigma_{*,\rm e}$ for PG\,2130+099
is notably higher than our quoted value, but the change does not
qualitatively affect our analysis. We use the corrected \sigstar
measurements ($\sigma_{*,\rm e}$) for the subsequent discussion and
analysis.

\subsection{Recalculation of AGN Virial Products}
There are now 30 reverberation-mapped AGNs with \sigstar
measurements. The virial products for most of these objects were
calculated from time lags that were determined with traditional cross
correlation methods (e.g., \citealt{Peterson04};
\citealt{Bentz09c}). However, \cite{Zu11} recently introduced a
different method to determine time lags that they called Stochastic
Process Estimation for AGN Reverberation (SPEAR). This method works as
follows: We assume all emission-line light curves are scaled and
shifted versions of the continuum light curve. The continuum is
modeled as an autoregressive process using a damped random walk (DRW)
model, which has shown to be a good statistical model of quasar
variability (\citealt{Gaskell87}; \citealt{Kelly09};
\citealt{Kozlowski10}; \citealt{MacLeod10}; \citealt{Zu12}). The
transfer function is modeled as a simple top-hat function. We fit the
continuum and emission-line light curves simultaneously, maximizing
the likelihood of the model using Bayesian Markov Chain Monte Carlo
iterations. The main advantage of the SPEAR method is that it treats
gaps in the temporal coverage of light curves in a statistically
self-consistent way, so the gaps in the data are filled in a
well-defined manner with well-defined uncertainties.

\cite{Zu11} re-measured the time lags in reverberation-mapped AGNs
with the SPEAR method and demonstrated its ability to recover accurate
time lags. The method has since been successfully used to improve RM
measurements (\citealt{Grier12b}; \citealt{Dietrich12}) and even
recover velocity-delay maps (\citealt{Grier13a}). Because many of the
light curves from the AGN \msigma sample, particularly high-luminosity
objects, have large gaps, we went through and re-determined the virial
products for the entire sample with updated \Hbeta \ time lags
recovered with the SPEAR method. Eighteen of the objects already have
\Hbeta \ time lags from \cite{Zu11}. For eight of the remaining
objects, we applied the same SPEAR method with the latest version of
the software called JAVELIN\footnote{Available at:
  http://www.astronomy.ohio-state.edu/$\sim$yingzu/codes.html\#javelin}
to calculate new time lags and virial products. Three of the remaining
AGNs have recently-published virial products calculated with the SPEAR
method, so we use the published virial products for 3C\,390.3 from
\cite{Dietrich12}, 3C\,120 from \cite{Grier12b}, and PG\,2130+099 from
\cite{Grier13a}. We do not have the light curve for Mrk 50, so for
this object we use the virial product from \cite{Barth11}. All
recalculated time lags, original line widths, and updated virial
products for each RM data set are given in Table
\ref{Table:tbl5}. For objects with just one measurement, we use that
virial product. For objects with multiple measurements, we adopt the
mean of the logarithm of the virial products. In Table
\ref{Table:tbl6} we show the adopted virial products and \sigstar
measurements for the entire sample.

In most cases, the updated virial products are very similar to the
previously-quoted values. The median fractional change in the virial
products due to the updated time lags is about 18\%, and the majority
of these changes are consistent to within the measurement errors. For
eight of the objects, the virial products changed somewhat
significantly. The largest increase was for 3C\,390.3, for which the
new virial product increased by factor of almost five over the
previous measurement. Most of these eight objects had light curves
with significant gaps at key locations in the light curves that likely
interfered with the time lag determinations. We do not see any
systematic increase or decrease in the fractional change as a function
of virial product (i.e., the virial products did not increase more in
objects with larger virial products, or vice versa, in general), but
we do note that the virial products of both of the highest-mass
objects, 3C\,390.3 and PG\,1426+015, both increased substantially from
previous measurements.

\section{Results and Discussion}
\subsection{Data Quality}

The objects for which we successfully measure \sigstar (PG\,1411+442,
PG\,1617+175, Mrk 509, and PG\,2130+099) are the four lowest-redshift
galaxies in our sample. Mrk 509 and PG\,2130+099 have the highest
$S/N$ host-galaxy spectra, with an average $S/N$ per pixel of $\sim$
250 for Mrk 509 and $\sim$ 190 for PG\,2130+099. These have the most
easily identifiable galaxy absorption features in our sample (see
Figure~3). The higher-luminosity quasars, PG\,1411+442
and PG\,1617+175, have lower $S/N$, specifically a $S/N$ of 100 per
pixel in PG\,1617+175 and 130 per pixel in PG\,1411+442, which is
lower than our anticipated $S/N$. Although we made several attempts to
fit and remove sky features, residual sky contamination and telluric
absorption lines remain in the subtracted spectra, which makes the
velocity dispersion measurements more uncertain and contributes to the
lower $S/N$ of the spectra. We also see stronger quasar emission
features which we were unable to eliminate entirely from the
host-galaxy spectra.

There seems to be three main factors that compromise the quality of
our host galaxy spectra. First, we were unable to satisfactorily
remove the sky emission from the spectra in all eight targets. This
caused a significant decrease in $S/N$ in all of our
spectra. Secondly, in three of our objects, PG\,0026+129,
PG\,1126+023, and PG\,0052+251, many of the strong stellar absorption
features that allow us to measure \sigstar were redshifted out of the
$H$-band. This limited us to very few absorption lines, and these few
remaining lines fell in regions with severe telluric contamination. As
such, we did not detect any absorption and were unable to measure
\sigstar in these objects. Third, and possibly most importantly, in
these four objects the quasar contamination becomes strong enough to
overwhelm the host galaxy flux despite our long integrations and
attempts to optimize the extraction radius. To quantify the amount of
quasar contamination remaining in the original extraction annuli of
these spectra, we estimated the ratio of the quasar flux to the host
flux, both within our extraction radius, for the case of
PG\,0026+129. We based this calculation on measurements of the PSF
magnitude, host galaxy magnitude, and host galaxy Sersic index
reported by \cite{Veilleux09} from their analysis of $HST$ NICMOS
$H$-band images. We integrate the Sersic profile over the extraction
annulus used in our study (for PG\,0026+129, we used an inner radius
of 0.2\arcsec and an outer radius of 0.6\arcsec, corresponding to 0.62
and 1.86 kpc, respectively) to estimate the amount of host flux inside
our aperture. Given the previously discussed findings of
\cite{Christou10}, we assume that half of the PSF light falls inside
the annulus and find that the PSF flux inside the extraction annulus
is a factor of 4.5 times the amount of host flux inside the extraction
annulus. We expect similar, possibly even more, contamination in the
other targets for which we were unsuccessful, and thus this
contamination limits our ability to explore the hosts of quasars at
the high end of the luminosity distribution.

While we appear to have reached the limit of the NIFS+ALTAIR system
for these measurements, AO systems continue to move towards
diffraction-limited resolution with high Strehl ratios. These advances
may lead to successful measurements with similar exposure times. The
future availability of the $James \ Webb\ Space\ Telescope$\ ($JWST$)
may also lead to successful attempts at \sigstar measurements in
high-luminosity quasars. $JWST$ is currently expected to launch in
2018 and will be equipped with an IFU spectrograph of sufficient
resolution, and the major problems of sky and telluric contamination,
which were prohibitive for our higher-redshift targets, will be
completely eliminated in space. Observing from space will allow us to
see the whole spectrum continuously, so we will not be limited to
specific redshift windows, and will also include the \caii \ triplet
region. $JWST$ will offer a compact, stable PSF, and will have more
sensitivity than our current equipment and thus offers much promise
for future efforts to measure \sigstar in high-luminosity,
high-redshift AGNs.

\subsection{The Faber-Jackson Relation}
It is also possible that the host galaxies of the four quasars in
which we were unable to measure \sigstar were simply fainter galaxies
with lower $\sigma_*$. If this is the case, excluding them in our
subsequent examination of the \msigma relation could result in a bias
in the distribution at the high-\mbh end. As a separate check on the
expected velocity dispersions of these hosts (as well as the rest of
the sample), we place the entire RM sample on the \cite{Faber76}
relation (hereafter the Faber-Jackson relation), which is the
correlation between \sigstar and the absolute magnitude or luminosity
of the host bulge. Some of the galaxies had already-published absolute
$V$-band magnitudes from \cite{Bentz09b}, and we used host galaxy
decompositions by \cite{Bentz09a} and \cite{Bentz13} to determine the
bulge magnitudes in the rest (except Mrk 50, which was not included in
these studies). To compare our sample with a previous determination of
the Faber-Jackson relation by \cite{Jiang11}, we convert our
magnitudes to the $I$ band. The $HST$ observations \citep{Bentz09a,
  Bentz13} were taken using the ACS F550M filter and the WFPC2 F547M
filter, while the images used by \cite{Jiang11} were taken with the
WFPC2 F814W filter. We use the IRAF package $synphot$ with bulge
templates from \cite{Kinney96} to determine the $m_{F550M} -
m_{F814W}$ and $m_{F547M} - m_{F814W}$ colors for each galaxy type and
use this to transform our bulge magnitudes to the magnitude in the
F814W filter, which is extremely close to the $I$ band.

\begin{figure}
\figurenum{4}
\begin{center}
\epsscale{1.2}
\plotone{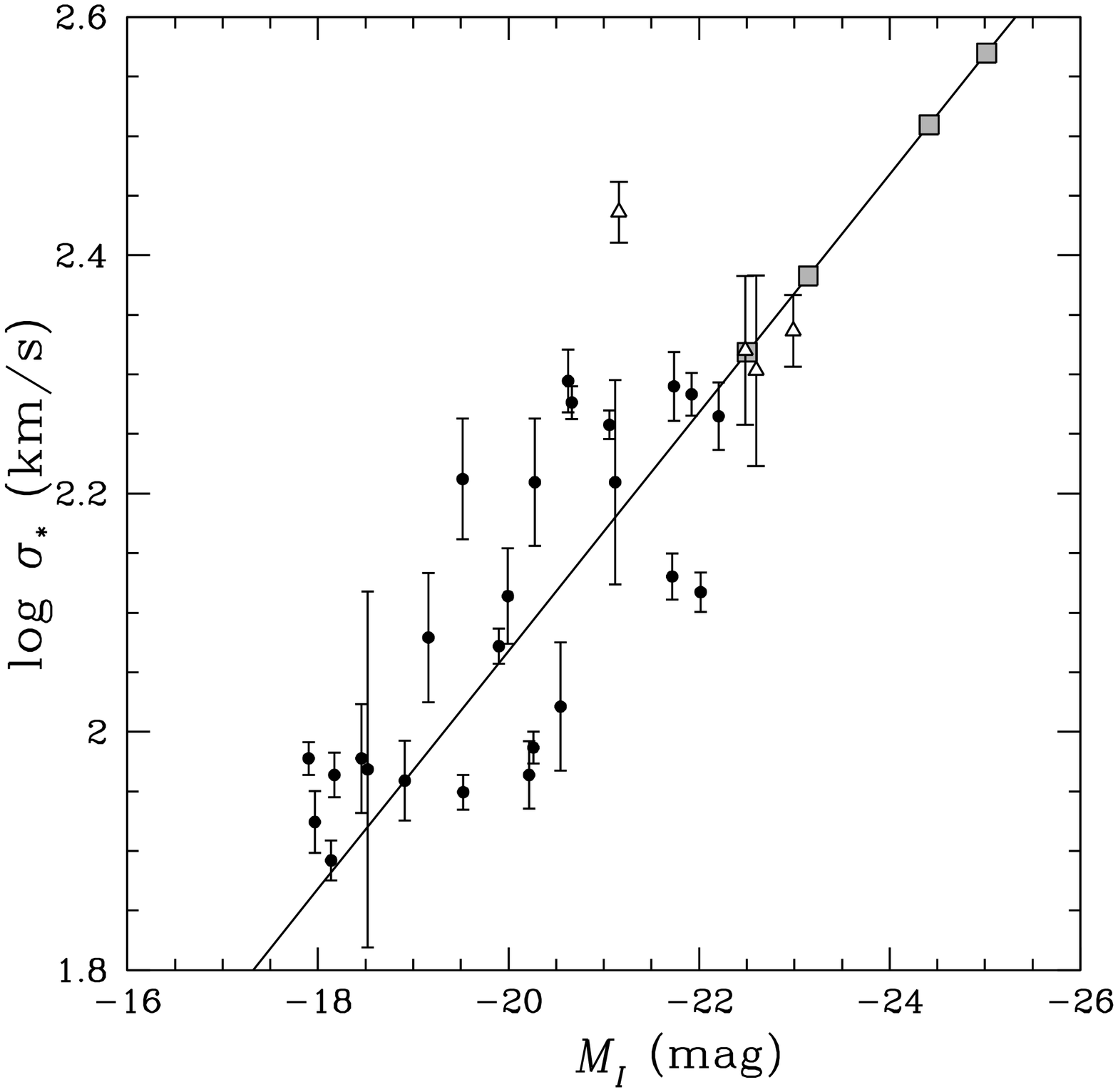}
\caption{The RM sample on the Faber-Jackson relation. The solid black
  line shows the relation measured by \citealt{Jiang11} with the
  \citealt{Gultekin09} galaxy sample. Filled black circles show the RM
  sample with secure \sigstar measurements. Open triangles show the
  locations of objects with \mbh $> 10^8$ \Msun, and gray filled
  squares show the expected locations of the four targets in our study
  for which we were unable to obtain \sigstar measurements, based on
  the relation from \citealt{Jiang11}. }
\label{fig:fabjack}
\end{center}
\end{figure}

Figure~4 shows our RM sample on the Faber-Jackson
relation from \cite{Jiang11}, which was calculated with the sample
from \cite{Gultekin09}. The RM sample as a whole appears to follow
this relation, with some scatter. We use the same relation measured by
\cite{Jiang11} with the sample from \cite{Gultekin09} to calculate the
expected \sigstar values for the four objects in which we were unable
to measure $\sigma_*$. As Figure~4 shows, the hosts of
the four targets in wich we were unable to measure \sigstar are some
of the most luminous in the sample, and thus the predicted \sigstar
values are quite high. As such, we conclude that our inability to
measure \sigstar in these objects is most likely due to the high AGN
luminosities and high redshifts of the systems and not due to
systematically fainter host galaxies in these targets. It is important
again to note that measuring bulge properties in AGN, particularly at
the high-luminosity end of the distribution, is very difficult, and
thus the bulge luminosities themselves are subject to large
uncertainties.

\subsection{The Virial Factor \boldmath{$\langle f \rangle$} and the
  {\boldmath \msigma} Relation}
Because we have updated the virial products in the AGN \msigma sample
and added a few objects at the high-luminosity end of the
distribution, we also present an updated measurement of the average
virial factor $\langle f \rangle$ used to calibrate the AGN \mbh
scale. In order to measure $\langle f \rangle$, we assume that the AGN
\msigma relation follows the same slope as quiescent
galaxies. However, previous studies have found that AGNs appear to
follow a slightly shallower relation than quiescent galaxies (e.g.,
\citealt{Woo10}; \citealt{Graham11}; \citealt{Park12};
\citealt{Woo13}). We measure the slope of the relation between the
virial product, $M_{\rm vir}$, and $\sigma_*$:
\begin{equation}
\rm {log} \ \it{M}_{\rm vir} = \alpha + \beta \ \rm{log}\ \left( \frac{\sigma_*}{200 \ km s^{-1}} \right)
\end{equation} 
with our updated AGN sample. We use the traditional forward regression
for our calculation: We consider \sigstar as the independent variable
and \mbh the dependent variable. Using the FITEXY algorithm
(\citealt{Press92}) to fit the AGN relation, we obtain
$\beta~=~5.04~\pm~0.19$. This slope is steeper than found in previous
studies (e.g., \citealt{Woo13}, who report $\beta$ = 3.46 $\pm$ 0.61),
and is slightly flatter than but consistent with the most recent
measurement of the slope in quiescent galaxies of 5.31 $\pm$ 0.33
(\citealt{Woo13}). We then use FITEXY to determine the $\langle f
\rangle$ necessary to place the AGN sample on the same \msigma
relation as quiescent galaxies. Using our updated AGN sample, we
obtain log~$\langle f \rangle$~=~0.63~$\pm$~0.11, corresponding to
$\langle f \rangle$~=~4.31~$\pm$~1.05. This number is slightly lower
than, but consistent with, the recent values found by \cite{Park12}
and \cite{Woo13} and is in closer agreement with that found by
\cite{Graham11} when using a forward regression analysis. However,
\cite{Graham11} also raised the issue of a potential sample selection
bias in the \msigma relation --- specifically, there are no \sigstar
measurements in objects with \mbh $< 10^6$ \Msun, causing a bias
against low-\mbh systems. They use an inverse regression approach to
avoid this selection bias. We prefer a forward regression because the
\msigma relation is typically used to determine \mbh in galaxies with
measured $\sigma_*$.

\begin{figure}
\figurenum{5}
\begin{center}
\epsscale{1.2}
\plotone{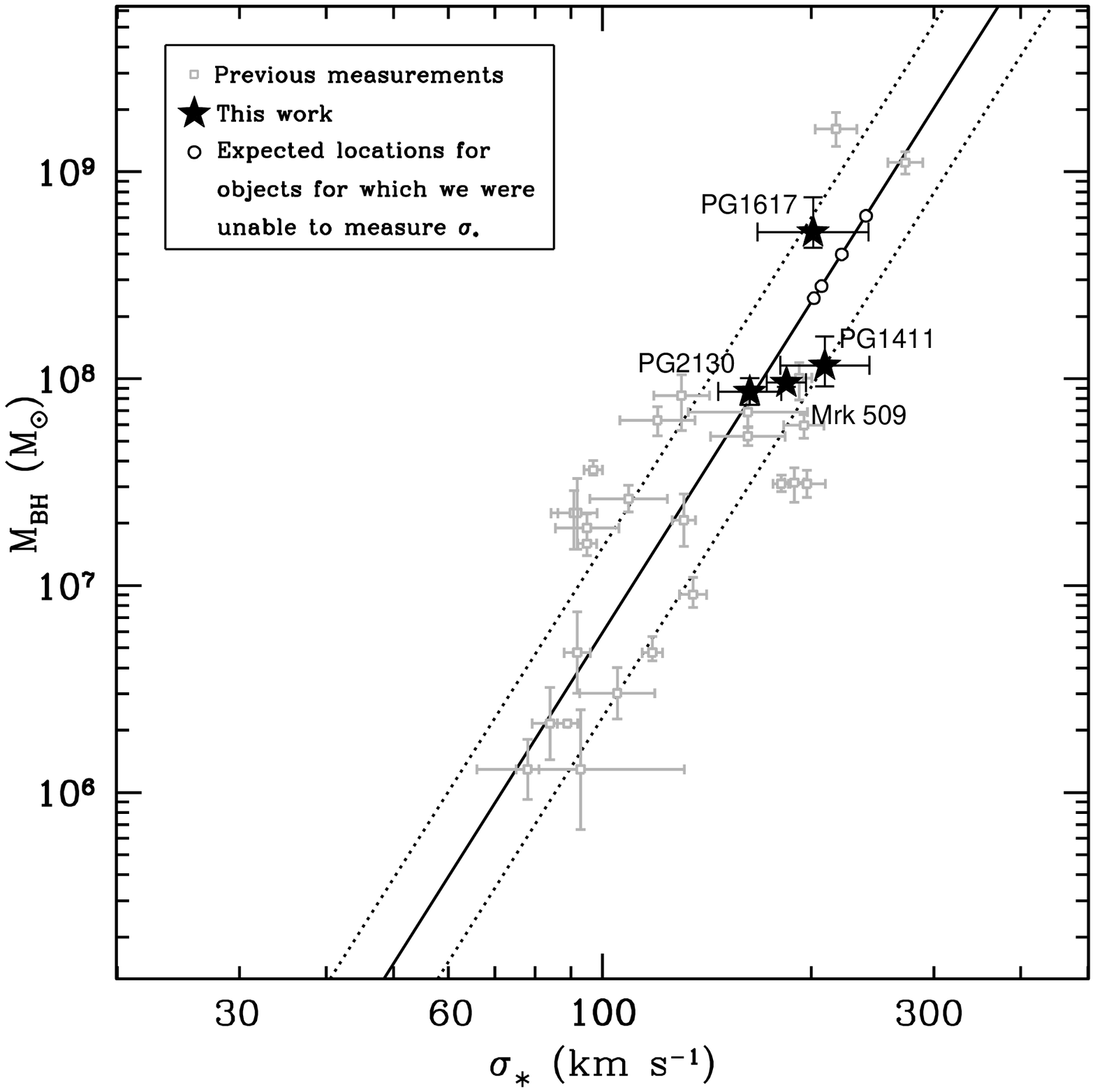}
\caption{The \msigma relation. The gray open squares are AGNs with
  previous measurements. The sample is composed of the compilation by
  \citealt{Woo10} and updated by \citealt{Park12} and \citealt{Woo13},
  with additional updates as described in the text (see Table
  \ref{Table:tbl6}). All \mbh were calculated with our measurement of
  $f$=4.31. The solid black stars show our new \sigstar
  measurements. The solid black line shows the most recent measurement
  of the \msigma relation in quiescent galaxies from \citealt{Woo13},
  and the dotted lines show the intrinsic scatter of the quiescent
  galaxies measured by \citealt{Woo13}. The open circles denote the
  expected locations of the objects for which we were unable to
  measure $\sigma_{*}$ based the \msigma relation.  }
\label{fig:msigma}
\end{center}
\end{figure}

We use our new value of $\langle f \rangle$ = 4.31 to transform the
virial products of the AGN sample to $M_{\rm BH}$. We place our four
objects with successful \sigstar measurements on the \msigma relation
in Figure~5, with the rest of the AGN sample shown for
comparison. We also show the most recent quiescent \msigma relation
from \cite{Woo13}. We see that all four of our objects lie within the
expected scatter of the relation, and see no evidence for an offset in
the objects at the high end of the relation. Because we were unable to
obtain \sigstar measurements for our highest-\mbh objects, we fall
short of our original goal of densely populating the very high end of
the \msigma relation, though we do increase the sample of objects with
10$^8$ \ \Msun $< $ \mbh $< 10^9 \ $\Msun. As noted in previous
studies, the AGN sample is still biased towards objects with
relatively lower masses than the majority of the quiescent galaxies
with dynamical mass measurements.

\begin{figure}
\figurenum{6}
\begin{center}
\epsscale{1.2}
\plotone{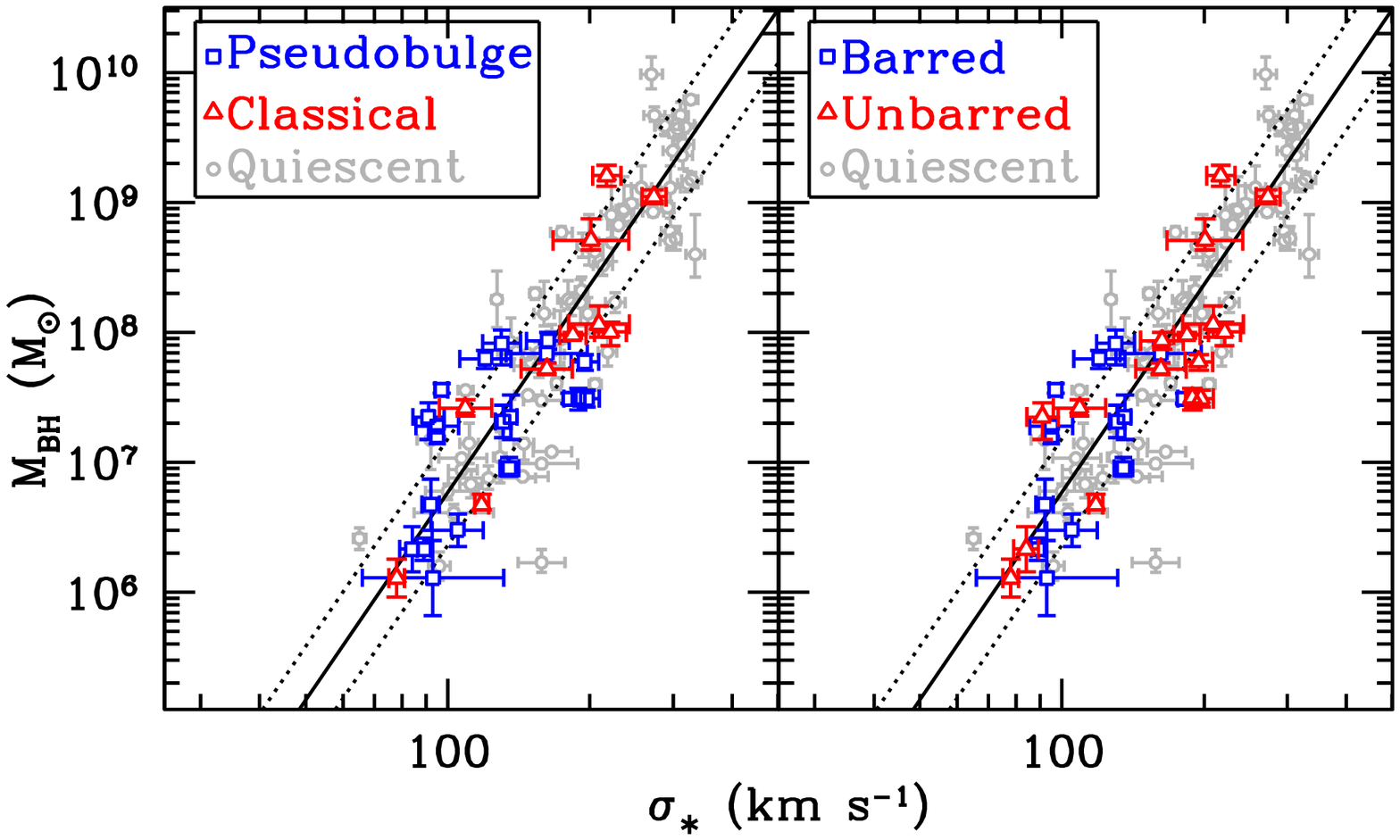}
\caption{The $M_{\rm BH}$-\sigstar relation for the same updated AGN
  sample shown in Figure~5. The quiescent galaxy sample
  of 72 galaxies from \cite{McConnell13}, with updates from
  \cite{Woo13}, is shown in gray for comparison. The solid black line
  shows the most recent measurement of the \msigma relation in these
  quiescent galaxies from \citealt{Woo13}, and the dotted lines show
  their measured intrinsic scatter in the quiescent relation. The left
  panel shows the AGN sample divided by the type of bulge (pseudobulge
  or classical), and the right panel shows the AGN sample sorted by
  the presence of a bar. All \mbh for the AGN sample were calculated
  with our measurement of $f$=4.31. }
\label{fig:msigmamorph}
\end{center}
\end{figure}

A morphological or environmental dependence of the \msigma relation
could have important implications on the use of a single mean $f$
factor to transform the virial product, $M_{\rm vir}$, into \mbh in
AGNs. To test whether or not the AGN sample shows any
morphology-dependent effects, we divided our AGN sample into different
groups to see if there was any visible offset or change in slope in
the \msigma relation. In the left panel of Figure~6, we divide the AGN
sample into two groups: those suspected of hosting a pseudobulge, and
those thought to host a classical spheroid at their centers. In the
right panel, we divide the sample based on whether or not a central
bar is observed. We classified the galaxies using the host-galaxy
decompositions from \cite{Bentz09a} (see their Table 4).  Following
\cite{Kormendy04}, the galaxies were inspected for the presence of the
following pseudobulge indicators: flattened bulge morphology, the
presence of a nuclear bar, a low ($n < 2$) Sersic index, and the
presence of copious dust and star formation in the nucleus without any
signature of an ongoing merger. None of these indicators by themselves
constitute sufficient evidence that a bulge is a
pseudobulge. Therefore, we assumed that galaxies with most of these
indicators host a pseudobulge, and galaxies with few or none of these
indicators host a classical bulge. Our classifications are listed in
Table \ref{Table:tbl6}. It should be noted that host galaxy
classifications are particularly difficult to make in high-luminosity
quasar hosts, and some galaxies in our sample show signs of disturbed
morphologies, so our classifications are somewhat uncertain.

We do not see any signs of an offset in either the galaxies hosting
pseudobulges or the galaxies with central bars, unlike several studies
that were previously noted in our introduction. If we determine
$\langle f \rangle$ with our samples of barred and unbarred galaxies
separately, we obtain a result similar to that of
\cite{Graham11}. Namely, the galaxies with bars in them yield a lower
value ($\langle f \rangle$~=~3.07~$\pm$~1.00) than the galaxies
without central bars ($\langle f \rangle
$~=~5.92~$\pm$~1.80). However, the barred sample lacks any objects
with \mbh $> 10^8$ \Msun and similarly, the unbarred sample has only
three objects with \mbh below 10$^7$ \Msun. Observations of the
megamaser sample by \cite{Greene10} show that about half of their host
galaxies (with \mbh $< 10^8$ \Msun) fall below the \msigma relation,
which again demonstrates that different samples yield different
results. In addition, \cite{Woo10, Woo13} and this work have
demonstrated that extending the sample to cover the entire \mbh range
can significantly impact the measured slopes and mean virial factor,
and as such we conclude the difference seen in our samples is not
statistically significant due to the systematic uncertainties
associated with the lack of a substantial dynamic range in $M_{\rm
  BH}$.

\section{Summary}
We measured \sigstar in four quasars at the high-mass end of the AGN
\msigma distribution. The measurements of Mrk 509 and PG\,1411+442 are
the first ever \sigstar measurements for these targets, and the
measurements for PG\,1617+175 and PG\,2130+099 are updates with
improved precision. We were unable to measure \sigstar in the spectra
of our highest-redshift objects due to substantial contamination of
the host-galaxy spectra by the quasar nucleus and poor sky line
subtraction. Future measurements of \sigstar for high-luminosity QSOs
will require improved AO technology, substantially longer integration
times, or $JWST$. We also updated the virial products for the AGN
\msigma sample with recalculated time lags from the SPEAR method of
\cite{Zu11} and recalculated the mean virial factor $\langle f
\rangle$ used to calibrate the AGN \mbh scale. We obtained log
$\langle f \rangle$ = 0.63 $\pm$ 0.11, corresponding to $\langle f
\rangle$~=~4.31 $\pm$ 1.05. This is consistent with previous results
based on forward regression methods. With our new \sigstar
measurements, all four of our objects fall within the expected scatter
of the quiescent \msigma relation. We find no evidence in AGNs for a
morphology-based deviation from the standard quiescent \msigma
relation.

\acknowledgments The authors would like to thank Jenny Greene and
Alister Graham for their helpful suggestions that improved this
work. C.J.G. would like to thank Gisella De Rosa for help with some of
the code used in this analysis. B.M.P., R.W.P., and C.J.G. gratefully
acknowledge the support of the National Science Foundation through
grant AST-1008882 to The Ohio State University. C.J.G. is also
supported by a Presidential Fellowship at The Ohio State
University. P.M. is grateful for support from the sabbatical visitor
program at the North American ALMA Science Center (NAASC) at NRAO and
the hospitality of both the NAASC and the University of Virginia while
this work was completed. This research has made use of the NASA/IPAC
Extragalactic Database (NED) which is operated by the Jet Propulsion
Laboratory, California Institute of Technology, under contract with
the National Aeronautics and Space Administration. This work is based
on observations obtained at the Gemini Observatory, which is operated
by the Association of Universities for Research in Astronomy, Inc.,
under a cooperative agreement with the NSF on behalf of the Gemini
partnership: the National Science Foundation (United States), the
National Research Council (Canada), CONICYT (Chile), the Australian
Research Council (Australia), Minist\'{e}rio da Ci\^{e}ncia,
Tecnologia e Inova\c{c}\~{a}o (Brazil) and Ministerio de Ciencia,
Tecnolog\'{i}a e Innovaci\'{o}n Productiva (Argentina).

\clearpage


\begin{deluxetable}{lccc} 
\tabletypesize{\scriptsize}
\setlength{\tabcolsep}{0.04in}
\tablecaption{Quasar Properties}
\tablewidth{0pt} 
\tablehead{ 
\colhead{} & 
\colhead{RA} & 
\colhead{DEC} & 
\colhead{$z$ }  \\  
\colhead{Galaxy} & 
\colhead{(J2000)} & 
\colhead{(J2000)} & 
\colhead{(NED)} 
} 
\startdata
PG\,0026+129    &	00 29 13.6      & 	+13 16 03     	&	0.142	 \\
PG\,0052+251    &	00 54 25.1      & 	+25 25 38     	&	0.154	 \\
PG\,1226+023    &	12 29 06.7      & 	+02 03 09     	&	0.158	 \\
PG\,1411+442    &	14 13 48.3      & 	+44 00 14     	&	0.089	 \\
PG\,1617+175    &	16 20 11.3      & 	+17 24 28     	&	0.112	 \\
PG\,1700+518    &	17 01 24.8      & 	+51 49 20     	&	0.292	 \\
Mrk 509         &	20 44 09.7      & 	-10 43 25    	&	0.034	 \\
PG\,2130+099    &	21 32 27.8      & 	+10 08 19     	&	0.063	
\enddata
\label{Table:tbl1}
\end{deluxetable} 

\begin{deluxetable}{lc} 
\tabletypesize{\scriptsize}
\setlength{\tabcolsep}{0.04in} 
\tablecaption{Most Prominent Stellar Absorption Features} 
\tablewidth{0pt}

\tablehead{ 
\colhead{Feature} & 
\colhead{Rest-frame}  \\ 
\colhead{} & 
\colhead{Wavelength ($\mu$m)} 
} 
\startdata
Mg I      &  1.4880   \\
Mg I      &  1.5030   \\
CO(3-0)   &  1.5580   \\
CO(4-1)   &  1.5780   \\
Si I      &  1.5890   \\
CO(5-2)   &  1.5980   \\
CO(6-3)   &  1.6190   \\
CO(8-5)   &  1.6610   \\
CO(9-6)   &  1.6840   \\
CO(10-7)  &  1.7060   \\

\enddata
\label{Table:tbl3}
\end{deluxetable} 

\begin{deluxetable}{lccccr} 
\tabletypesize{\scriptsize}
\setlength{\tabcolsep}{0.04in}
\tablecaption{Observations}
\tablewidth{0pt} 
\tablehead{ 
\colhead{} & 
\colhead{Observing } & 
\colhead{On-source} & 
\colhead{} &  \\ 
\colhead{Target} & 
\colhead{Semester} & 
\colhead{Integration time} &
\colhead{Band} \\
\colhead{} & 
\colhead{} & 
\colhead{(hours)} &
\colhead{} 
} 
\startdata
PG\,0026+129  	& 2008B	&	1.33	& H  \\
PG\,0052+251  	& 2010B	&	1.33	& H  \\
PG\,1226+023  	& 2010A	&	0.83	& H  \\
PG\,1411+442  	& 2010A	&	1.33	& H  \\
PG\,1617+175  	& 2010A	&	2.50	& H  \\
PG\,1700+518  	& 2010A	&	3.00	& K  \\ 
Mrk 509       	& 2008B &       2.00	& H  \\
PG\,2130+099  	& 2010B	&      	1.00	& H  

\enddata
\label{Table:tbl2}
\end{deluxetable} 


\begin{deluxetable}{lcccccccc}
\tabletypesize{\scriptsize}
\setlength{\tabcolsep}{0.04in}
\tablecaption{Extraction Windows and Measurements}
\tablewidth{0pt} 
\tablehead{ 
\colhead{Galaxy} & 
\colhead{$R_{\rm inner}$ } & 
\colhead{$R_{\rm outer}$ } & 
\colhead{Stellar} & 
\colhead{\sigstar} &
\colhead{Best-fit } &
\colhead{} &
\colhead{} &
\colhead{$\sigma_{*,\rm e}$} \\ 
\colhead{} & 
\colhead{(arcsec)} & 
\colhead{(arcsec)} & 
\colhead{Template} & 
\colhead{(\kms)}  &
\colhead{$\chi^2$ } &
\colhead{$DOF$} & 
\colhead{$\chi^2/DOF$} &
\colhead{(\kms)} 
}
\startdata
PG\,0026+129    &   0.2	   & 0.6  & \nodata  & \nodata          &\nodata  & \nodata & \nodata & \nodata  \\
PG\,0052+251    &   0.2    & 0.7  & \nodata  & \nodata          &\nodata  & \nodata & \nodata & \nodata  \\
PG\,1226+023  	&   0.4    & 1.4  & \nodata  & \nodata          &\nodata  & \nodata & \nodata & \nodata  \\
PG\,1411+442    &   0.2    & 1.6  & K5\,III  & 216 $\pm$ 31     & 858.80  & 914    & 0.940    & 209 $\pm$ 30    \\
PG\,1617+175    &   0.3    & 1.3  & M5Ia     & 201 $\pm$ 37     & 491.33  & 649    & 0.757    & 201 $\pm$ 37     \\
PG\,1700+518    &   0.4    & 1.1  & \nodata  & \nodata          &\nodata  & \nodata & \nodata & \nodata  \\
Mrk 509         &   0.3    & 1.7  & K5\,III  & 189 $\pm$ 12     & 1046.17 & 1134   & 0.923    & 184 $\pm$ 12    \\
PG\,2130+099    &   0.2	   & 1.2  & K5\,III  & 147 $\pm$ 17     & 548.70  & 914    & 0.600    & 163 $\pm$ 19    
\enddata
\tablecomments{$R_{\rm inner}$ and $R_{\rm outer}$ correspond to the
  inner and outer radii of the circular extraction annulus for each
  object. $\chi^2$ and $\chi^2/DOF$ are reported for the best fits
  with the K5\,III stellar template for PG\,1411+442, Mrk 509 and
  PG\,2130+099. For the case of PG\,1617, the M5Ia template was
  used. $R_{\rm inner}$ and $R_{\rm outer}$ correspond to the inner
  and outer radii of the extraction annulus. $\sigma_{*,\rm e}$ was
  calculated with the formula for E and S0 galaxies from
  \cite{Jorgensen95}.}
\label{Table:tbl4}
\end{deluxetable} 

\begin{deluxetable}{lccccc}
\tabletypesize{\scriptsize}
\setlength{\tabcolsep}{0.04in}
\tablecaption{Reverberation Measurements and Virial Products}
\tablewidth{0pt} 
\tablehead{ 
\colhead{Galaxy} & 
\colhead{$\tau_{\rm H\beta}$\tablenotemark{a}} &
\colhead{$\tau_{\rm H\beta}$} &
\colhead{$\sigma_{\rm line}$(rms)} &
\colhead{$\sigma_{\rm line}$(rms)} &
\colhead{$M_{\rm vir}$}  \\
\colhead{} & 
\colhead{(days)} &
\colhead{Reference} &
\colhead{(\kms)} &
\colhead{Reference}  &
\colhead{(10$^6$ \Msun)}  
}
\startdata
3C\,120	         &  27.2$^{+1.1}_{-1.1}$	& 	1	&	1514 $\pm$ 65	&	1	& 12.2$^{+1.2}_{-1.2}$       \\
3C\,390.3        &  44.3$^{+3.0}_{-3.3}$	&       2	&	5455 $\pm$ 278	&	2	& 260$^{+36}_{-23}$	       \\
Ark 120	         &  35.7$^{+6.7}_{-9.2}$	&	3	&	1959 $\pm$ 109	&	7	& 26.7$^{+5.8}_{-7.5}$    \\
Ark 120          &  29.7$^{+3.3}_{-5.9}$	&	3	&	1884 $\pm$ 48 	&	7	& 20.6$^{+2.5}_{-4.2}$  \\
Arp 151	         &  3.6$^{+0.7}_{-0.2}$	&	4	&	1252 $\pm$ 46	&	8, 9	& 1.1$^{+0.2}_{-0.1}$     \\
Mrk 50	         &  10.4$^{+0.8}_{-0.9}$	&	5	&	1740 $\pm$ 101	&	5	& 6.2$^{+0.9}_{-0.9}$	\\
Mrk 79	         &  25.5$^{+2.9}_{-14.4}$  &	3	&	2137 $\pm$ 375	&	7	& 22.7$^{+8.4}_{-15.1}$	\\
Mrk 79           &  30.9$^{+1.4}_{-2.1}$   &	3	&	1683 $\pm$ 72 	&	7	& 17.1$^{+1.7}_{-1.9}$    \\
Mrk 79           &  17.2$^{+7.3}_{-2.2}$   &	3	&	1854 $\pm$ 72 	&	7	& 11.5$^{+5.0}_{-1.7}$     \\
Mrk 79           &  43.6$^{+1.7}_{-0.8}$   &	3	&	1883 $\pm$ 246	&	7	& 30.1$^{+7.9}_{-7.9}$     \\
Mrk 110	         &  25.3$^{+2.3}_{-13.1}$  &	3	&	1196 $\pm$ 141	&	7	& 7.1$^{+1.8}_{-4.0}$     \\
Mrk 110	         &  33.9$^{+6.1}_{-5.3}$   &	3	&	1115 $\pm$ 103	&	7	& 8.2$^{+2.1}_{-2.0}$     \\
Mrk 110	         &  21.5$^{+2.2}_{-2.1}$   &	3	&	755  $\pm$ 29 	&	7	& 2.4$^{+0.3} _{-0.3}$    \\
Mrk 202	         &  3.5$^{+0.1}_{-0.1}$	&	4	&	659  $\pm$ 65	&	8, 9	& 0.30$^{+0.06}_{-0.06}$     \\ 
Mrk 279	         &  18.3$^{+1.2}_{-1.1}$	&	3	&	1420 $\pm$ 96	&	7	& 7.2$^{+1.1}_{-1.1}$     \\
Mrk 509	         &  69.9$^{+0.3}_{-0.3}$	&	3	&	1276 $\pm$ 28	&	7	& 22.2$^{+1.0}_{-1.0}$	\\
Mrk 590	         &  19.0$^{+1.9}_{-2.6}$	&	3	&	789  $\pm$ 74	&	7	& 2.3$^{+0.5}_{-0.5}$	\\
Mrk 590          &  19.5$^{+2.0}_{-4.0}$	&	3	&	1935 $\pm$ 52	&	7	& 14.2$^{+1.7}_{-3.0}$     \\
Mrk 590	         &  32.6$^{+3.5}_{-8.8}$	&	3	&	1251 $\pm$ 72	&	7	& 9.9$^{+1.6}_{-2.9}$      \\
Mrk 590	         &  30.9$^{+2.5}_{-2.4}$	&	3	&	1201 $\pm$ 130	&	7	& 8.7$^{+2.0}_{-2.0}$      \\
Mrk 817	         &  20.9$^{+2.3}_{-2.3}$	&	3	&	1392 $\pm$ 78	&	10	& 7.9$^{+1.2}_{-1.2}$	\\
Mrk 817	         &  17.2$^{+1.9}_{-2.7}$	&	3	&	1971 $\pm$ 96	&	10	& 13.0$^{+1.9}_{-2.4}$     \\
Mrk 817	         &  35.9$^{+4.8}_{-5.8}$	&	3	&	1729 $\pm$ 158	&	10	& 20.9$^{+4.7}_{-5.1}$     \\
Mrk 817	         &  10.8$^{+1.5}_{-1.0}$	&	3	&	3150 $\pm$ 295	&	10	& 20.9$^{+4.9}_{-4.4}$   	\\
Mrk 1310         &  4.2$^{+0.9}_{-0.1}$	&	4	&	755  $\pm$ 138	&	8, 9	& 0.5$^{+0.2}_{-0.2}$	\\
NGC\,3227        &  10.6$^{+6.1}_{-6.1}$	&       3	&	1925 $\pm$ 124	&	10	& 7.7$^{+4.5}_{-4.5}$	\\
NGC\,3227        &  4.4$^{+0.3}_{-0.5}$	&       3	&	2018 $\pm$ 174	&	10	& 3.5$^{+0.7}_{-0.7}$ 	\\
NGC\,3516        &  14.6$^{+1.4}_{-1.1}$	&	3	&	1591 $\pm$ 10	&	10	& 7.2$^{+0.7}_{-0.6}$	\\
NGC\,3783        &  7.3$^{+0.3}_{-0.7}$	&	3	&	1753 $\pm$ 141	&	7	& 4.4$^{+0.7}_{-0.8}$	\\
NGC\,4051        &  2.5$^{+0.1}_{-0.1}$	&	3	&	1034 $\pm$ 41	&	10	& 0.5$^{+0.1}_{-0.1}$	\\
\enddata
\tablenotetext{a}{Time lags are all given in the rest frame.}

\tablecomments{This table has been abbreviated and can be downloaded from the arXiv in full with the article 
source files. The published version of the paper will include the full table. References:
  1. \citealt{Grier12b}; 2. \citealt{Dietrich12}; 3. \citealt{Zu11};
  4. This work;  5. \citealt{Barth11}; 6. \citealt{Grier13a}; 7. \citealt{Peterson04};
  8. \citealt{Bentz09b}; 9. \citealt{Park12}.
  10. \citealt{Denney10} 11. \citealt{Bentz06a};
  12. \citealt{Denney06}}
\label{Table:tbl5}
\end{deluxetable} 
\begin{deluxetable}{lccccc}
\tabletypesize{\scriptsize}
\setlength{\tabcolsep}{0.04in}
\tablecaption{The AGN \msigma Sample}
\tablewidth{0pt} 
\tablehead{ 
\colhead{Galaxy} & 
\colhead{} & 
\colhead{$M_{\rm vir}$\tablenotemark{b}} & 
\colhead{$M_{\rm BH}$\tablenotemark{c}} & 
\colhead{\sigstar} &
\colhead{\sigstar} \\
\colhead{} & 
\colhead{Classification\tablenotemark{a}}  & 
\colhead{(10$^6$ \Msun)} & 
\colhead{(10$^6$ \Msun)} & 
\colhead{(\kms)} &
\colhead{Reference} 
}
\startdata
3C\,120	        &	Classical	       & 12.2$^{+1.2}_{-1.2}$   &    52.6$^{+5.2}_{-5.2}$                      &  162 $\pm$ 20	&    1 \\
3C\,390.3	&	Classical	       & 260$^{+36}_{-23}$     &    1120.6$^{+138.0}_{-142.2}$                           &  273 $\pm$ 16&    2 \\
Ark 120	        &	Classical	       & 23.4$^{+4.0}_{-5.7}$   &    100.9$^{+17.2}_{-24.6}$                        &  192 $\pm$ 8	&    3 \\
Arp 151	        &	Classical	       & 1.1$^{+0.2}_{-0.1}$    &    4.7$^{+0.9}_{-0.4}$                    &  118 $\pm$ 4	&    4 \\
Mrk 50	        &	Classical	       & 6.2$^{+0.9}_{-0.9}$    &    26.3$^{+3.9}_{-3.9}$                     &  109 $\pm$ 14	&    5 \\
Mrk 79	        &	Barred Pseudobulge     & 19.2$^{+4.5}_{-7.4}$   &    82.8$^{+19.4}_{-31.9}$                        & 130 $\pm$ 12	&    2\\
Mrk 110	        &	Pseudobulge	       & 5.2$^{+1.3}_{-2.1}$    &    22.4$^{+5.6}_{-9.1}$                     &  91 $\pm$ 7	&    6\\
Mrk 202	        &	Classical	       & 0.3$^{+0.1}_{-0.1}$    &    1.3$^{+0.4}_{-0.4}$                    &  78 $\pm$ 3	&    4\\
Mrk 279	        &	Pseudobulge	       & 7.2$ ^{+1.1}_{-1.1}$   &    31.0$^{+4.7}_{-4.7}$                     &  197 $\pm$ 12	&    2\\
Mrk 509	        &	Classical	       & 22.2$^{+1.0}_{-1.0}$   &    95.7$^{+4.3}_{-4.3}$                     &  184 $\pm$ 12	&    7\\
Mrk 590	        &	Pseudobulge	       & 7.3$^{+1.2}_{-1.6}$    &    31.5$^{+5.2}_{-6.9}$                     &  189 $\pm$ 6	&    2\\
Mrk 817	        &	Barred Pseudobulge     & 14.6$^{+2.2}_{-2.5}$   &    62.9$^{+9.5}_{-10.8}$                      &  120 $\pm$ 15	&    2\\
Mrk 1310	&	Pseudobulge	       & 0.47$^{+0.20}_{-0.17}$  &   2.2$^{+0.9}_{-0.9}$                     &  84 $\pm$ 5	&    4\\
NGC\,3227	&	Barred Pseudobulge     & 5.2$^{+2.0}_{-2.1}$    &    22.4$^{+8.6}_{-9.1}$                     &  92 $\pm$ 6	&    3\\
NGC\,3516	&	Barred Pseudobulge     & 7.2$^{+0.7}_{-0.6}$    &    31.0$^{+3.0}_{-2.6}$                     &  181 $\pm$ 5	&    2\\
NGC\,3783	&	Barred Pseudobulge     & 4.4$^{+0.7}_{-0.8}$    &    19.0$^{+3.0}_{-3.4}$                     &  95 $\pm$ 10	&    8\\
NGC\,4051	&	Barred Pseudobulge     & 0.5$^{+0.1}_{-0.1}$    &    2.2$^{+0.2}_{-0.4}$                    &  89 $\pm$ 3	&    2\\
NGC\,4151	&	Barred Pseudobulge     & 8.4$^{+0.9}_{-0.5}$    &    36.2$^{+3.9}_{-2.2}$                     &  97 $\pm$ 3	&    2\\
NGC\,4253	&	Barred Pseudobulge     & 0.3$^{+0.2}_{-0.2}$    &    1.3$^{+0.9}_{-0.9}$                    &  93 $\pm$ 32	&    4\\
NGC\,4593	&	Barred Pseudobulge     & 2.1$^{+0.4}_{-0.3}$    &    9.1$^{+1.7}_{-1.3}$                    &  135 $\pm$ 6	&    2\\
NGC\,4748	&	Barred Pseudobulge     & 0.7$^{+0.2}_{-0.2}$    &    3.0$^{+0.9}_{-0.9}$                    &  105 $\pm$ 13	&    4\\
NGC\,5548	&	Pseudobulge	       & 13.8$^{+1.7}_{-2.0}$   &    59.5$^{+7.3}_{-8.6}$                     &  195 $\pm$ 13	&    4\\
NGC\,6814	&	Barred Pseudobulge     & 3.7$^{+0.5}_{-0.5}$    &    15.9$^{+2.2}_{-2.2}$                     &   95 $\pm$ 3	&    4\\
NGC\,7469	&	Barred Pseudobulge     & 4.8$^{+1.4}_{-1.4}$    &    20.7$^{+6.0}_{-6.0}$                     &  131 $\pm$ 5	&    2\\
PG\,1229+204	&	Barred Pseudobulge     & 16.0$^{+2.7}_{-2.6}$   &    69.0$^{+11.6}_{-11.2}$                   &  162 $\pm$ 32	&    9\\
PG\,1411+442	&	Classical              & 26.9$^{+8.7}_{-6.3}$   &    115.9$^{+37.5}_{-27.2}$                    &  209 $\pm$ 30	&    7 \\
PG\,1426+015	&	Classical	       & 373.3$^{+68.7}_{-71.6}$&    1609$^{+296}_{-309}$                     &  217 $\pm$ 15	&    10\\
PG\,1617+175	&	Classical	       & 118.6$^{+45.8}_{-20.6}$ &  511.2$^{+197.4}_{-88.8}$                  &  201 $\pm$ 37 &    7\\
PG\,2130+099	&	Pseudobulge	       & 20.1$^{+3.0}_{-3.0}$   &   86.6$^{+12.9}_{-12.9}$                    &  163 $\pm$ 19	&    7\\
SBS\,1116+583A	&	Barred Pseudobulge     & 1.1$^{+0.5}_{-0.5}$    &   4.7$^{+2.2}_{-2.2}$                     &   92 $\pm$ 4	&    4

\enddata
\tablenotetext{a}{Classifications were made using the host galaxy
  decompositions of \cite{Bentz09a} and the critera are discussed in
  the text.}  

\tablenotetext{b}{For objects with multiple RM measurements, the
  adopted virial product is the average of the logarithm of the
  different virial products.}
  \tablenotetext{c}{\mbh were computed
  using $f$ = 4.31. } \tablecomments{Velocity Dispersion References:
  1. \citealt{Nelson95}; 2. \citealt{Nelson04}; 3. \citealt{Woo13};
  4. \citealt{Woo10}; 5. \citealt{Barth11}; 6. \citealt{Ferrarese01};
  7. This work; 8. \citealt{Onken04}; 9. \citealt{Dasyra07};
  10. \citealt{Watson08}. }
\label{Table:tbl6}
\end{deluxetable} 

\end{document}